\documentclass[preprint,nofootinbib,floatfix,a4paper,prd,superscriptaddress]{revtex4-1}
\pdfoutput=1
\usepackage{hyperref}

\usepackage{graphicx}
\usepackage[utf8]{inputenc} 
\usepackage{amsmath,amssymb}
\usepackage{slashed}
\usepackage{epstopdf}
\usepackage{geometry}

\usepackage{rotating}

\usepackage{xcolor}
\usepackage{tabularx}
\newcolumntype{C}{>{\centering\arraybackslash}X} 
\def\be{\begin{equation}}
\def\ee{\end{equation}}
\def\bea{\begin{eqnarray}}
\def\eea{\end{eqnarray}}

\newcommand{\oh}{\ensuremath{\Omega_{D}h^2}}
\newcommand{\sigmav}{\ensuremath{\langle\sigma v\rangle}}
\newcommand{\sigsip}{\ensuremath{\sigma^{\rm{SI}}_p}}

\newcommand{\mev}{\ensuremath{\text{~MeV}}}
\newcommand{\gev}{\ensuremath{\text{~GeV}}}
\newcommand{\tev}{\ensuremath{\text{~TeV}}}

\def  \bcen   {\begin{center}}
\def  \ecen   {\end{center}}
\def  \beq    {\begin{equation}}
\def  \eeq    {\end{equation}}
\def  \bpm    {\begin{pmatrix}}
\def  \epm    {\end{pmatrix}}
\def  \beqa   {\begin{eqnarray}}
\def  \eeqa   {\end{eqnarray}}
\def\bea{\begin{eqnarray}}
\def\eea{\end{eqnarray}}
\def  \nn     {\nonumber }

\def\la   {\lambda}

\def\nn{\nonumber}
\def\lee { \left( }
\def\rii { \right) }
\def\lan   {\langle}
\def\ran   {\rangle}

\def\De {\Delta}

\def\to {\rightarrow}

\begin{document}

\title{Complex Scalar Dark Matter in G2HDM}

\author{Chuan-Ren Chen}
\affiliation{\small Department of Physics, National Taiwan Normal University, Taipei 116, Taiwan}

\author{Yu-Xiang Lin}
\affiliation{\small Department of Physics, National Taiwan Normal University, Taipei 116, Taiwan}

\author{Chrisna Setyo Nugroho}
\affiliation{\small Institute of Physics, Academia Sinica, Nangang, Taipei 11529, Taiwan}
\affiliation{\small Physics Division, National Center for Theoretical Sciences, Hsinchu 300, Taiwan}

\author{Raymundo Ramos}
\affiliation{\small Institute of Physics, Academia Sinica, Nangang, Taipei 11529, Taiwan}

\author{Yue-Lin Sming Tsai}
\affiliation{\small Institute of Physics, Academia Sinica, Nangang, Taipei 11529, Taiwan}
\affiliation{\small Key Laboratory of Dark Matter and Space Astronomy, Purple
Mountain Observatory, Chinese
Academy of Sciences, Nanjing 210008, China}

\author{Tzu-Chiang Yuan}
\affiliation{\small Institute of Physics, Academia Sinica, Nangang, Taipei 11529, Taiwan}

\date{\today}

\begin{abstract}

The complex scalar dark matter (DM) candidate in the gauged two Higgs doublet model 
(G2HDM), stabilized by a peculiar hidden parity ($h$-parity), 
is studied in detail. We explore the parameter space for the DM candidate by taking into account 
the most recent DM constraints from various experiments, in particular, the PLANCK relic density
measurement and the current DM direct detection limit from XENON1T.  We
separate our analysis in three possible compositions for the mixing of the
complex scalar.  We first constrain our parameter space with the vacuum
stability and perturbative unitarity conditions for the scalar potential, 
LHC Higgs measurements, plus Drell-Yan and electroweak precision test 
constraints on the gauge sector. 
We find that DM dominated by composition of the inert doublet scalar is completely excluded 
by further combining the previous constraints with both the latest results from PLANCK and XENON1T.
We also demonstrate that the remaining parameter space with two other DM compositions 
can be further tested by indirect detection like the future CTA gamma-ray telescope.  
 
\end{abstract}

\maketitle

\section{Introduction}

Dark Matter (DM) has become one of the most discussed topics 
in cosmology, astrophysics and particle physics.  However, besides the indirect evidence from 
the power spectrum from the cosmic microwave background radiation (CMB)
and the galaxy rotational curves which provide strong hints for the need of DM, 
current experiments for DM direct detection, indirect detection 
and collider searches still show no clue for the nature of DM.
Currently, the best description of the early history of the Universe is given
by the $\Lambda$CDM model which assumes the presence of dark energy and cold
DM in additional to the ordinary matter. The leading hypothesis is that this cold DM
is comprised of weakly interacting massive particle (WIMP) that was thermally
produced just like the other Standard Model (SM) particles in the early universe.  
It is well known that the most compelling
feature of WIMP DM is that, after freeze-out whence the DM reaction rates fell 
behind the Hubble expansion rate of the universe,
it is possible to achieve the correct relic abundance with
an electroweak sized annihilation cross section with a WIMP mass of a
few hundreds GeV to a few TeV.

On the collider phenomenology side, we now know that all the major
decay modes of the SM 125 GeV Higgs discovered on the fourth of July 2012 at the Large Hadron Collider (LHC) 
have been observed except the $Z \gamma$ and $\mu^+\mu^-$ modes.
So far all the experimental results agree with the SM predictions within 10 $\sim$ 15\%. 
Nevertheless there are still some rooms for new physics. 
A particular class of models that extends simply the scalar sector of the SM to address 
new physics is quite popular.
The most well known example is the general two Higgs doublet model (2HDM),
which has several variants and resulted in very rich phenomenology. 
For a review of 2HDM and its phenomenology, see for example~\cite{Branco:2011iw,Logan:2014jla}.
One of the interesting variants of 2HDM is to impose a ${\cal Z}_2$ symmetry in the model 
so that the second Higgs doublet is $\mathcal Z_2$-odd and then can be a DM candidate.
This is the inert Higgs doublet model (IHDM)~\cite{Deshpande:1977rw} 
and many detailed phenomenological studies~\cite{LopezHonorez:2006gr, Arina:2009um, Nezri:2009jd, Miao:2010rg,
Gustafsson:2012aj, Arhrib:2012ia, Arhrib:2014pva,Swiezewska:2012eh, Arhrib:2013ela,
Goudelis:2013uca, Krawczyk:2013jta, Krawczyk:2013pea, Ilnicka:2015jba,
Diaz:2015pyv, Modak:2015uda, Kephart:2015oaa,Queiroz:2015utg, Garcia-Cely:2015khw,
Hashemi:2016wup, Poulose:2016lvz, Alves:2016bib, Datta:2016nfz,
Belyaev:2016lok, Belyaev:2018ext} had been performed over the years. 
Furthermore, the idea that this ${\mathcal Z}_2$ symmetry emerges accidentally 
in a renormalizable gauged two Higgs doublet model
(G2HDM) has been explored recently in~\cite{Huang:2015wts}. 
In G2HDM the two Higgs doublets are grouped together in an irreducible doublet 
representation of an extra non-Abelian $SU(2)_H$ gauge group. 
Besides the new hidden $SU(2)_H$, the SM gauge group is
extended by including a new $U(1)_X$ symmetry.  The scalar sector is further
extended by including a new $SU(2)_H$ doublet and a triplet, both singlets
under the SM gauge group.

Although any electrically neutral ${\cal Z}_2$-odd particle in G2HDM can be considered a DM
candidate, such as $W^\prime$ or heavy new neutrinos, in this work we choose to
concentrate on the phenomenology of the complex scalar DM candidate. The
reason is mainly that we want to present this model as a viable alternative to
IHDM and as such our setup is focused in providing a light, neutral and ${\cal
Z}_2$-odd complex scalar. Another reason is practicality, given that our setup
is complicated enough to distinguish at least 3 main types of DM candidates
coming only from the scalar sector due to mixing effects. We will study in detail the differences,
similarities and results of these three possibilities. Phenomenologically, we
expect all of them to communicate with the SM through Higgs portal
type interactions~\cite{Davoudiasl:2004be,Ham:2004cf}. 
However, we will demonstrate that the SM $Z$ boson as well as its heavier siblings in G2HDM
will also play non-negligible roles as mediators in various DM processes in relic density, direct and indirect 
detection, especially for the inert doublet-like DM case.

In the past, some collider phenomenology of G2HDM has been
studied~\cite{Huang:2015rkj, Huang:2017bto, Chen:2018wjl}. It was determined
that Drell-Yan type signals may help detect the G2HDM $Z^\prime$ in the high
luminosity upgrade of the LHC~\cite{Huang:2017bto} and that enhancement of pair production of Higgs
boson in the LHC is moderate compared to the SM~\cite{Chen:2018wjl}. In a recent study it has been
determined that G2HDM has a viable scalar sector parameter
space~\cite{Arhrib:2018sbz}, compatible with vacuum stability and perturbative
unitarity conditions, as well as Higgs phenomenology constraints from the LHC.  The gauge sector is constrained by electroweak
precision tests (EWPT)~\cite{Huang:2019obt}, setting limits on the masses of the new
gauge bosons and the gauge sector parameter space.  It is precisely this two
recent studies on the scalar and gauge sectors constraints ({\bf SGSC}) that
we will take as starting point for our study, thus ensuring that the final
constrained parameter space is consistent with previous studies and that our
result has a stronger relevance.

This article is organized as follows.
In Sec.~\ref{sec:model} we briefly recall some salient features of the G2HDM
model, in particular the scalar potential and mass spectra. 
In Sec.~\ref{sec:HiddenP} we point out after spontaneous symmetry breaking 
there exists an accidental discrete $\mathcal Z_2$ symmetry in the whole Lagrangian of G2HDM.
We classify all the particles in the model according to whether they are even or odd under this 
discrete symmetry, dubbed as $h$-parity.
This residual symmetry forbids the lightest particle in the hidden sector 
to decay and hence it, if electrically neutral, may be a cold DM candidate in the model.
We discuss further the compositions of complex scalar dark matter that is relevant in this work.
In Sec.~\ref{sec:constraints}, we discuss the DM constraints included in our
analysis and how they are affected by our setup in more general terms.
We describe how relic density ({\bf RD}), direct detection ({\bf
DD}) and indirect detection ({\bf ID}) measurements constrain each of the
three different compositions of complex scalar DM considered. 
We also discuss the collider searches of DM from the mono-jet plus missing energy search 
and invisible Higgs decay. 
In Sec.~\ref{sec:result}, after a brief description of the methodology used in our numerical
analysis we present the results of our analysis.
Finally, in Sec.~\ref{sec:summary} we summarize our findings and conclude,
including a brief comment on future detectability. 
Some Feynman rules that are most relevant to the 
processes discussed in this work are collected in the Appendix~\ref{sec:appendix}.

\vfill

\section{The G2HDM Model}
\label{sec:model}

\subsection{Matter Content} 
\begin{table}[htbp!]
	\begin{tabular}{|c|c|}
		\hline
		Matter Fields & $SU(3)_C \times SU(2)_L \times SU(2)_H \times U(1)_Y \times U(1)_X$ \\
		\hline \hline
		$H=\left( H_1\,, \; H_2 \right)^{\rm T}$ & (1, 2, 2, 1/2, 1) \\
		$\Delta_H$ & (1, 1, 3, 0, 0) \\
		$\Phi_H$ & (1, 1, 2, 0, $1$) \\
		\hline\hline
		$Q_L=\left( u_L\,, \; d_L \right)^{\rm T}$ & (3, 2, 1, 1/6, 0)\\
		$U_R=\left( u_R\,, \; u^H_R \right)^{\rm T}$ & (3, 1, 2, 2/3, 1) \\
		$D_R=\left( d^H_R\,, \; d_R \right)^{\rm T}$ & (3, 1, 2, $-1/3$, $-1$) \\
		\hline
		$L_L=\left( \nu_L\,, \; e_L \right)^{\rm T}$ & (1, 2, 1, $-1/2$, 0) \\
		$N_R=\left( \nu_R\,, \; \nu^H_R \right)^{\rm T}$ & (1, 1, 2, 0, $1$) \\
		$E_R=\left( e^H_R\,, \; e_R \right)^{\rm T}$ & (1, 1, 2, $-1$, $-1$) \\
		\hline
		$\nu_L^H$ & (1, 1, 1, 0, 0) \\
		$e_L^H$ & (1, 1, 1, $-1$, 0) \\
		\hline
		$u_L^H$ & (3, 1, 1, 2/3, 0) \\
		$d_L^H$ & (3, 1, 1, $-1/3$, 0) \\
		\hline
	\end{tabular}
	\caption{Matter contents and their quantum number assignments in G2HDM. 
	}
	\label{tab:quantumnos}
\end{table}

The gauge symmetry group of G2HDM expands the SM gauge group $SU(2)_L \times U(1)_Y$ 
by adding a hidden sector of $SU(2)_H \times U(1)_X$.  In the scalar sector we have the two
$SU(2)_L$ Higgs doublets $H_1$ and $H_2$ both paired into an $SU(2)_H$ doublet
$H$.  The two scalar SM singlets, $\Delta_H$ and $\Phi_H$, have been put into
the triplet and doublet representations of $SU(2)_H$, respectively. 
In order to construct gauge invariant Yukawa couplings, new
right-handed heavy fermions have been added as $SU(2)_H$ companions of the SM
right-handed fermions, pairing both of them together into $SU(2)_H$ doublets,
but remaining $SU(2)_L$ singlets.  Anomaly cancellation requires further that we add
two pairs of left-handed heavy leptons and two pairs of left-handed heavy
quarks for each family, all of them are singlets under both $SU(2)$ groups and
under $U(1)_X$. 
We note that the $SU(2)_H$ is not the same as the $SU(2)_R$ in 
left-right symmetric models~\cite{Mohapatra:1979ia,Keung:1983uu}.
The $W^{\prime (p,m)}$ in G2HDM does not carry electric charge, while the $W^{\prime \pm}$ in 
left-right symmetric models does. Thus we use the superscripts $p$ and $m$ to label them, instead of $+$ and $-$.
We also note that non-sterile right-handed neutrinos $\nu_{l R}$s introduced in the mirror fermion models of electroweak scale right-handed neutrinos~\cite{Hung:2006ap,Hung:2017voe,Hung:2015hra} 
are in a different manner.
In the mirror fermion models, $\nu_{l R}$s are grouped with mirror charged leptons $l^M_R$s to form $SU(2)_L$ doublets.
Here in G2HDM, they are grouped with new heavy right-handed neutrinos $\nu^H_{l R}$ 
to form $SU(2)_H$ doublets instead.
For other related ideas extending the 2HDM with extra gauge symmetries to address flavor problem, 
dark matter and neutrino masses, see for example~\cite{Ko:2012hd,Campos:2017dgc,Camargo:2018klg,Camargo:2018uzw,Camargo:2019ukv,Cogollo:2019mbd}.
The matter contents of the G2HDM model and their respective quantum numbers
are listed in Table~\ref{tab:quantumnos}.

\subsection{Scalar Potential and Constraints}
\label{subsec:pottheorcons}

\subsubsection*{Scalar Potential}

For this work we will be using the scalar potential from
Ref.~\cite{Arhrib:2018sbz}, that extends the original potential of
Ref.~\cite{Huang:2015wts} by adding two new terms with couplings $\lambda'_H$
and $\lambda'_{H\Phi}$. 
The most general scalar potential that respects the G2HDM symmetries can be
divided into 4 different parts
\begin{equation}
V_T = V (H) + V (\Phi_H ) + V ( \De_H ) + V_{\rm mix} \left( H, \Delta_H, \Phi_H \right) \; .
\label{eq:higgs_pots} 
\end{equation}
The first term $V(H)$ in Eq.~\eqref{eq:higgs_pots} consists of the two Higgs doublets $H_1$ and $H_2$ only and is given by
\begin{align}
\label{VH1H2}
V(H)
= {}& \mu^2_H   \left( H^{\alpha i}  H_{\alpha i} \right)
+  \la_H \left( H^{\alpha i}  H_{\alpha i} \right)^2  
+ \frac{1}{2} \la'_H \epsilon_{\alpha \beta} \epsilon^{\gamma \delta}
\left( H^{ \alpha i}  H_{\gamma  i} \right)  \left( H^{ \beta j}  H_{\delta j} \right) \nn \\
= {}& \mu^2_H   \left( H^\dag_1 H_1 + H^\dag_2 H_2 \right) 
+ \la_H   \left( H^\dag_1 H_1 + H^\dag_2 H_2 \right)^2  \nn \\
{}& \;\;\;\;\;\;\;\;\;\;\;\;\;\;\;\;\;\;\;\;\;\;\;\;\;\;\;\;\;\;\; + \la'_H \left( - H^\dag_1 H_1 H^\dag_2 H_2 
+ H^\dag_1 H_2 H^\dag_2 H_1 \right)  \; , 
\end{align}
where greek and latin letters refer to $SU(2)_H$ and $SU(2)_L$ indices respectively, 
both of which run from 1 to 2, and we use the notation $H^{\alpha i} = H^*_{\alpha i}$. 
From the second line of Eq.~\eqref{VH1H2}, one can clearly see that 
$V(H)$ has the discrete $\mathcal Z_2$ symmetry of $H_1 \to H_1$ and $H_2 \to - H_2$. 
Since $V(H)$ contains all the renormalizable terms constructed solely from $H_1$ and $H_2$,
this discrete symmetry is automatically present. Recall that in general 2HDM,
one needs to impose this discrete symmetry to avoid unwanted terms that may lead to 
flavor changing neutral current in the Higgs-Yukawa interactions at the tree level.
The second term $V ( \Phi_H )$ is for the $SU(2)_H$ doublet $\Phi_H$ only and given by
\begin{align}
\label{VPhi}
V ( \Phi_H )
= {}& \mu^2_{\Phi}   \Phi_H^\dag \Phi_H  + \la_\Phi \lee \Phi_H^\dag \Phi_H  \rii^2 \nn \\
= {}& \mu^2_{\Phi} \lee \Phi^*_1\Phi_1 + \Phi^*_2\Phi_2 \rii +  \la_\Phi \lee \Phi^*_1\Phi_1 + \Phi^*_2\Phi_2 \rii^2 \; , 
\end{align}
where $\Phi_H = (\Phi_1\,, \; \Phi_2)^{\rm T}$.
The third term is for the $SU(2)_H$ triplet $\De_H$ and is given by
\begin{align}
\label{VDeltas}
V ( \De_H ) = {}& - \mu^2_{\De} {\rm Tr} \lee \De^2_H  \rii  \;  + \la_\De \lee {\rm Tr} \lee \De^2_H  \rii \rii^2 \nn \\
= {}& - \mu^2_{\De} \lee \frac{1}{2} \De^2_3 + \De_p \De_m  \rii +  \la_{\De} \lee \frac{1}{2} \De^2_3 + \De_p \De_m  \rii^2 \; , 
\end{align}
where
\begin{align}
\De_H=
\begin{pmatrix}
	\De_3/2   &  \De_p / \sqrt{2}  \\
	\De_m / \sqrt{2} & - \De_3/2   \\
\end{pmatrix} = \De_H^\dagger \; {\rm with}
\;\; \Delta_m = \left( \Delta_p \right)^* \; {\rm and} \; \left( \Delta_3 \right)^* = \Delta_3 \;   .
\end{align}
Furthermore, unlike other models with $SU(2)_L$ triplet Higgs, 
the off-diagonal components $\De_{p,m}$ do not carry electric charge. We use
the subscripts $p$ and $m$ to label them instead of $+$ and $-$, in the same
way as the new gauge bosons $W^{\prime (p,m)}$.
Finally, the last term $V_{\rm{mix}}$ consists all three scalars $H$, $\Phi_H$ and $\De_H$ 
\begin{align}
\label{VvMix}
V_{\rm{mix}} \left( H , \Delta_H, \Phi_H \right) = 
{}& + M_{H\De}  \lee H^\dag \De_H H \rii -  M_{\Phi\De}  \lee \Phi_H^\dag \De_H \Phi_H \rii  \nn \\
{}& + \la_{H\Phi} \lee H^\dag H  \rii  \lee \Phi_H^\dag \Phi_H \rii  
+ \la^\prime_{H\Phi} \lee H^\dag \Phi_H  \rii  \lee \Phi_H^\dag H \rii
\nn\\
{}& +  \la_{H\De} \lee H^\dag H  \rii    {\rm Tr} \lee \De^2_H  \rii
+ \la_{\Phi\De} \lee \Phi_H^\dag \Phi_H \rii {\rm Tr} \lee \De^2_H \rii  \; . 
\end{align}
Eq.~\eqref{VvMix} can be expanded further in terms of the component fields of $H$, $\Delta_H$ and $\Phi_H$ as follows
\begin{align}
V_{\rm{mix}} \left( H , \Delta_H, \Phi_H \right) =
{}& + M_{H\De} \lee \frac{1}{\sqrt{2}}H^\dag_1 H_2 \De_p  
+  \frac{1}{2} H^\dag_1 H_1\De_3 + \frac{1}{\sqrt{2}}  H^\dag_2 H_1 \De_m  
- \frac{1}{2} H^\dag_2 H_2 \De_3   \rii   \nn \\
{}& - M_{\Phi\De} \lee  \frac{1}{\sqrt{2}} \Phi^*_1 \Phi_2 \De_p  
+  \frac{1}{2} \Phi^*_1 \Phi_1\De_3 + \frac{1}{\sqrt{2}} \Phi^*_2 \Phi_1 \De_m  
- \frac{1}{2} \Phi^*_2 \Phi_2 \De_3   \rii  \nn \\
{}& +  \la_{H\Phi} \lee H^\dag_1 H_1 + H^\dag_2 H_2 \rii  \lee \Phi^*_1\Phi_1 + \Phi^*_2\Phi_2 \rii \nn\\
{}& +  \la^\prime_{H\Phi} \lee H^\dag_1 H_1 \Phi^*_1\Phi_1 + H^\dag_2 H_2  \Phi^*_2\Phi_2 
+ H^\dag_1 H_2 \Phi_2^*\Phi_1 + H^\dag_2 H_1  \Phi^*_1\Phi_2  \rii \nn\\
{}& + \la_{H\De} \lee H^\dag_1 H_1 + H^\dag_2 H_2 \rii   \lee \frac{1}{2} \De^2_3 + \De_p \De_m  \rii \nn\\
{}& + \la_{\Phi\De}  
\lee  \Phi^*_1\Phi_1 + \Phi^*_2\Phi_2 \rii  \lee \frac{1}{2} \De^2_3 + \De_p \De_m  \rii \; .
\label{eq:vmix}
\end{align}
Interestingly, $V_{\rm mix} \left( H , \Delta_H, \Phi_H \right)$ is invariant under the combined discrete symmetry
$H_1 \to H_1$, $H_2 \to - H_2$, $\Phi_1 \to - \Phi_1$, $\Phi_2 \to  \Phi_2$, 
$\De_3 \to  \De_3$, and $\De_{p,m} \to  - \De_{p,m}$.
Since the complete scalar potential $V_T$ consists of all renormalization terms 
constructed out of $H, \Delta_H$ and $\Phi_H$, this discrete symmetry 
can be viewed as an accidental one in the scalar sector of the model. 
In fact one can extend this discrete symmetry to the whole renormalizable Lagrangian of G2HDM. We will discuss this further
in Sec~\ref{sec:HiddenP}.

\subsubsection*{Spontaneous Symmetry Breaking (SSB)}
\label{subsubsec:symbreak}

The gauge symmetry of G2HDM is broken spontaneously by the vacuum expectation values (VEVs) 
of $\langle H_1 \rangle = (0, v/\sqrt 2)^{\rm T}$, 
$\langle \Phi_2 \rangle = v_\Phi/\sqrt 2$, and $\lan \De_3 \ran = - v_\De$. 
In Ref.~\cite{Huang:2015wts}, we demonstrated that $v_\De$ satisfies a cubic equation with all coefficients 
expressed in terms of the fundamental parameters in the scalar potential.
Solutions of $v_\De$ can be found either 
analytically or numerically and plug into the linear coupled equations for $v^2$ and $v_\Phi^2$ 
that can then be solved straightforwardly. Since $v_\De$ has three different roots in general, 
the correct one will be picked by minimum energy requirement. Thus, the symmetry breaking of 
$SU(2)_L \times U(1)_Y \times U(1)_X$ in G2HDM is induced or triggered by the triplet VEV $v_\De$ which breaks $SU(2)_H$.
Note that the sign of $\mu^2_\Delta$ is negative with respect to $\mu^2_\Phi$
and $\mu^2_H$. If $\mu^2_\De > 0$, $SU(2)_H$ is spontaneously broken by the VEV $v_\De \neq 0$.
After $SU(2)_H$ symmetry breaking is triggered, the vacuum alignment of
$\Phi_H$ is controlled by the quadratic terms for $\Phi_1$ and $\Phi_2$ given
by
\begin{eqnarray}
\label{Phi1massterms}
&&\mu^2_\Phi + \frac{1}{2} M_{\Phi\De} v_\De + \frac{1}{2} \lambda_{\Phi \De} v_\De^2 
+  \frac{1}{2} ( \lambda_{H\Phi} + \lambda^\prime_{H\Phi} ) 
v^2 \; , \\
&&\mu^2_\Phi - \frac{1}{2} M_{\Phi\De} v_\De + \frac{1}{2} \lambda_{\Phi \De} v_\De^2 
+  \frac{1}{2} \lambda_{H\Phi} 
v^2 \; , 
\label{Phi2massterms}
\end{eqnarray}
respectively. The parameters $M_{\Phi\Delta}$, $\lambda_{\Phi\Delta}$, $\lambda_{H\Phi}$ and
$\lambda'_{H\Phi}$ can be either positive or negative independent of the
sign of $\mu^2_\Phi$, meaning that Eqs.~\eqref{Phi1massterms} and
\eqref{Phi2massterms} can be positive and negative, respectively.  One can
achieve $\langle \Phi_1 \rangle = 0$ and $\langle \Phi_2 \rangle \neq 0$ 
by making judicious choices of the parameters.
Furthermore, $SU(2)_L$ symmetry breaking is
controlled by the quadratic terms of $H_1$ and $H_2$.  After expanding the
potential, the coefficients for them are
\begin{eqnarray}
\label{H1massterms}
&&\mu^2_H - \frac{1}{2} M_{H\De} v_\De + \frac{1}{2} \lambda_{H \De} v_\De^2 
+  \frac{1}{2} \lambda_{H\Phi} 
v_\Phi^2 \; , \\
&&\mu^2_H + \frac{1}{2} M_{H\De} v_\De + \frac{1}{2} \lambda_{H \De} v_\De^2 
+  \frac{1}{2} (   \lambda_{H\Phi} +  \lambda^\prime_{H\Phi} ) 
v_\Phi^2 \; , 
\label{H2massterms}
\end{eqnarray} 
respectively. Similar to the $\Phi_H$ case, the parameters involved can be
either positive or negative independently of the sign of $\mu^2_H$, but this
time Eqs.~\eqref{H1massterms} and \eqref{H2massterms} would be negative and
positive, respectively. In this case, the gauge symmetry $SU(2)_L$ is broken
by the vacuum alignment $\langle H_1 \rangle \neq 0$ and $\langle H_2
\rangle = 0$. 
In our numerical scan for the parameter space in Sec.~\ref{sec:result}, we 
will search for parameters such that Eqs.~\eqref{Phi1massterms} and \eqref{H2massterms} are positive while 
Eqs.~\eqref{Phi2massterms} and \eqref{H1massterms} are negative in order to achieve the desired vacuum alignment.
The $\mu^2_H$, $\mu^2_\Phi$ and $\mu^2_\Delta$ parameters will be
fixed using the VEV equations (Eqs.~(18) to (20) of
Ref.~\cite{Huang:2015wts}).
Since $H_1$, $\Phi_{2}$ and $\De_3$ are all even under $\mathcal Z_2$, 
the discrete symmetry is not broken by their VEVs. Therefore 
the $\mathcal Z_2$-odd $H_2$ can become a DM candidate as long as
it is lighter than all other $\mathcal Z_2$-odd particles in the model.

\subsubsection*{Theoretical and Phenomenological Constraints on the Scalar Potential}

We will begin our analysis by considering the conditions determined in
Ref.~\cite{Arhrib:2018sbz} for the scalar sector parameter space. Namely, we
want to start with a parameter space that leaves the scalar potential bounded
from below, with couplings that remain within perturbative unitarity ranges
and make sure that we can actually achieve a sufficiently SM-like Higgs with a
$\sim$125~GeV mass and that can pass the limits set by the LHC. 

While for the minimum of the potential one checks the quadratic terms, to
ensure that the scalar potential is bounded from below for large-field values
one is mainly concerned with the quartic terms. In Ref.~\cite{Arhrib:2018sbz}
it was show that copositivity criteria~\cite{Kannike:2012pe,Kannike:2016fmd,Klimenko:1984qx} 
is enough to find conditions for the
potential to be bounded from below and have a stable vacuum. The copositive conditions are
\begin{equation}
\widetilde \lambda_H (\eta) \geq 0 \; , \;\;\; \lambda_\Phi \geq 0 \; , \;\;\; \lambda_\Delta \geq 0 \; ,
\label{coposA}
\end{equation}
\begin{eqnarray}
\label{coposB}
\Lambda_{H\Phi}(\xi , \eta) & \equiv & \widetilde \lambda_{H\Phi}(\xi) + 2 \sqrt{\widetilde \lambda_H (\eta) \lambda_\Phi}  \geq  0 \; , \nonumber \\
\Lambda_{H\Delta} (\eta) & \equiv & \lambda_{H\Delta} + 2 \sqrt{\widetilde \lambda_H (\eta) \lambda_\Delta}  \geq  0 \; , \\
\Lambda_{\Phi \Delta} & \equiv & \lambda_{\Phi \Delta} + 2 \sqrt{\lambda_\Phi \lambda_\Delta}  \geq  0 \; , \nonumber
\end{eqnarray}
\begin{eqnarray}
\Lambda_{H\Phi\Delta} (\xi , \eta) \equiv \sqrt{\widetilde \lambda_H (\eta) \lambda_\Phi \lambda_\Delta}
& + &  \frac{1}{2} \left(
	\widetilde \lambda_{H \Phi}(\xi) \sqrt{\lambda_\Delta}
	+ \lambda_{H \Delta} \sqrt{\lambda_\Phi}
	+ \lambda_{\Phi \Delta} \sqrt{\widetilde \lambda_H (\eta)}
\right) \nonumber \\
& & \;\;\; + \; \frac{1}{2} \sqrt {\Lambda_{H\Phi}(\xi , \eta) \Lambda_{H\Delta} (\eta) \Lambda_{\Phi \Delta} }
\geq 0 \; ,
\label{coposC}
\end{eqnarray}
where the parameters $\xi$ and $\eta$ can have any value in the ranges $0 \leq
\xi \leq 1$ and $-1/4 \leq \eta \leq 0$.

On the other hand we have to make sure that our parameter space remains within
perturbative limits. Again we look only at quartic couplings since 2$\to$2
scattering processes induced by cubic couplings are suppressed by their
propagators while quartic couplings are not. After checking all the possible
2$\to$2 scattering processes the final ranges allowed by perturbative
unitarity are
\begin{equation}
\begin{array}{cc}
\vert \lambda_i ( \mathcal M_1 ) \vert \leq 8 \pi \; , \;  \forall i = (1,\cdots , 10) \, ,& \\
\vert \lambda_H \vert  \leq  4 \pi \; , \vert \lambda_H^\prime \vert \leq 8 \sqrt 2 \pi \; ,
\vert 2 \lambda_H \pm \lambda^\prime_H \vert \leq 8 \pi \; , \vert  \lambda_\Phi \vert  \leq  4 \pi \; , \vert \lambda_\De \vert  \leq   4 \pi \; , &\\
\vert \lambda_{H\Phi} \vert  \leq  8 \pi \; ,
\vert \widetilde \lambda_{H\Phi} \vert = \vert \lambda_{H\Phi} + \lambda_{H\Phi}^\prime \vert \leq  8 \pi \; ,
\vert \lambda^\prime_{H\Phi} \vert  \leq  8 \sqrt 2 \pi \; , & \\
\vert \lambda_{H\Delta} \vert  \leq  8 \pi \; ,  \vert \lambda_{\Phi\Delta} \vert  \leq  8 \pi \; , &
\end{array}
\label{pertunitarity3}
\end{equation}
where
\begin{eqnarray}
&\lambda_1 = 2\lambda_H\, , \:\:\:\:\: \lambda_2 = 2\lambda_\Phi\, , \:\:\:\:\:
\lambda_3 = 2\lambda_\Delta\, , \:\:\:\:\: \lambda_{4,5} = 2\lambda_H \pm
\lambda^\prime_H \, ,\nonumber\\
&\lambda_{6,7} = \widetilde{\lambda}^+_H +
\lambda_\Phi \pm \sqrt{2\lambda^{\prime 2}_{H\Phi} +
(\widetilde{\lambda}^+_H - \lambda_\Phi)^2} \, ,
\end{eqnarray}
with $\widetilde{\lambda}^+_H \equiv  \lambda_H + \lambda_H^\prime/2$ and
$\lambda_{8,9,10}$ given by the three roots of the equation $\lambda^3 + a\lambda^2 +
b\lambda + c = 0$ with
\begin{align}
a = {}& - 5\lambda_\Delta - 6\lambda_\Phi - 10\lambda_H
+ \lambda^\prime_H \; , \nonumber \\
b = {}& - 6\lambda_{H\Delta}^2 - 3\lambda_{\Phi\Delta}^2
+ 5\lambda_\Delta(10\lambda_H - \lambda^\prime_H + 6\lambda_\Phi)
+ 6\lambda_\Phi(10\lambda_H - \lambda^\prime_H)
\nonumber \\
& - 8(\lambda_{H\Phi} + \lambda^\prime_{H\Phi}/2)^2 \; ,  \\
c  = {}& 36\lambda_\Phi\lambda_{H\Delta}^2
- 24\lambda_{H\Delta}\lambda_{\Phi\Delta}(\lambda_{H\Phi}+ \lambda^\prime_{H\Phi}/2)
+ 40\lambda_\Delta(\lambda_{H\Phi} + \lambda^\prime_{H\Phi}/2)^2
\nonumber \\
&
+ (3\lambda_{\Phi\Delta}^2
- 30\lambda_\Delta\lambda_\Phi)(10\lambda_H - \lambda^\prime_H)
\; . \nonumber
\end{align}

On the phenomenological side, we will require the presence of a SM Higgs with a mass of
$125.09 \pm 0.24$~GeV and a signal strength for the Higgs decay into
two photos of $\mu^{\gamma\gamma}_{ggH}=0.81^{+0.19}_{-0.18}$ as found by the
ATLAS experiment~\cite{Aaboud:2018xdt}. For more details about the theoretical
conditions described here, we encourage the interested reader to consult
Ref.~\cite{Arhrib:2018sbz}.

\subsection{Mass Spectra}
\label{sec:MassSpectrum}

\subsubsection*{Higgs-like  ($\mathcal Z_2$-even) Scalars}

Expanding the scalar potential in terms of the VEVs and taking the second derivatives with respect to the scalar fields, one can obtain the mass terms and the mixing terms of the scalar fields. The SM Higgs is extracted from the mixing of
three real scalars $h$, $\phi_2$ and $\delta_3$~\footnote{We follow the notations of~\cite{Huang:2015wts} shifting the scalar fields as:
$H_1 = \left( \begin{array}{c} G^+ \\ \frac{v + h}{\sqrt 2} + i G^0 \end{array} \right)$, 
$H_2 = \left( \begin{array}{c} H^+ \\ H^0_2 \end{array} \right)$, 
$\Phi_H = \left( \begin{array}{c} G^p_H \\ \frac{v_\Phi + \phi_2}{\sqrt 2} + i G^0_H \end{array} \right)$, 
and 
$\De_H = \left( 
\begin{array}{cc} 
\frac{-v_\De + \delta_3}{2} & \frac{\De_p}{\sqrt 2} \\ \frac{\De_m}{\sqrt2} & \frac{v_\De - \delta_3}{2}
\end{array} \right)$.}. 
The mixing matrix of these ${\cal Z}_2$-even neutral real scalars written in the
basis of ${\cal S}=\{h, \phi_2, \delta_3\}^{\rm T}$ is given by
\begin{align}
{\mathcal M}_0^2 =
\begin{pmatrix}
	2 \lambda_H v^2 & \lambda_{H\Phi} v v_\Phi 
	& \frac{v}{2} \left( M_{H\De} - 2 \lambda_{H \De} v_\De \right)  \\
	\lambda_{H\Phi} v v_\Phi
	& 
	2 \lambda_\Phi v_\Phi^2
	&  \frac{ v_\Phi}{2} \left( M_{\Phi\De} - 2 \lambda_{\Phi \De} v_\De \right) \\
	\frac{v}{2} \left( M_{H\De} - 2 \lambda_{H \De} v_\De \right)  & \frac{ v_\Phi}{2} \left( M_{\Phi\De} - 2 \lambda_{\Phi \De} v_\De \right) & \frac{1}{4 v_\De} \left( 8 \lambda_\De v_\De^3 + M_{H\Delta} v^2 + M_{\Phi \De} v_\Phi^2 \right)   
\end{pmatrix} \; .
\label{eq:scalarbosonmassmatrix}
\end{align}
The physical fields with definite mass can be obtained by doing the similarity transformation to this mixing matrix via orthogonal rotation matrix, ${\cal O}$, in such a way 
that
\begin{equation}
{\cal O}^{\rm T}\cdot {\mathcal M}_0^2 \cdot {\cal O} = {\rm Diag}(m^2_{h_1}, m^2_{h_2}, m^2_{h_3}) \; ,
\label{eq:OTM0sqO}
\end{equation}
where the masses of the fields are arranged in ascending manner $m_{h_1} \leq
m_{h_2} \leq m_{h_3}$. The interaction basis ${\cal S}$ and mass eigenstates
are related through the ${\cal O}$ mixing matrix via ${\cal S} = {\cal O}\cdot
\{h_1, h_2, h_3\}^{\rm T}$. In this setup, the 125 GeV Higgs
boson observed at the LHC is identified by the lightest mass eigenstate $h_{1}$. 

Other ${\cal Z}_2$-even scalars are the massless would be 
Goldstone bosons $G^{\pm,0}$ and $G^0_H$ which do not mix with other scalar fields. 
However they mix with the longitudinal components of the gauge fields and will be absorbed away.

\subsubsection*{Dark  ($\mathcal Z_2$-odd) Scalars}

The charged Higgs is sitting at the upper component of $H_2$ which
acquires mass from all three VEVs but it does not mix with other fields.
Since $H_2$ couples to all three multiplets $H_1$, $\Phi_H$ and $\Delta_H$,
after SSB it acquires tree mass terms one with each VEV given by
\begin{align} m^2_{H^\pm} &= M_{H \De} v_\De  - \frac{1}{2}\la^\prime_H v^2
+\frac{1}{2}\lambda^\prime_{H\Phi}v_\Phi^2 \;.  \label{chargedHiggsmass}
\end{align}

The complex fields $G^{p,m}_H$ , $H^{0(*)}_2$ and $\Delta_{p,m}$~\footnote{See previous footnote 
for the definitions of these complex scalars.} 
also acquire mass terms and
mix. In the basis of $\mathcal{G}=\{ G^p_H , H^{0*}_2, \Delta_p  \}^T$, the
squared mass matrix is given by
\begin{align}
{\mathcal M}_0^{\prime 2} =
\begin{pmatrix}
	M_{\Phi \Delta} v_\Delta +\frac{1}{2}\lambda^\prime_{H\Phi}v^2 & \frac{1}{2}\lambda^\prime_{H\Phi}  v v_\Phi & - \frac{1}{2} M_{\Phi \Delta} v_\Phi  \\
	\frac{1}{2}\lambda^\prime_{H\Phi} v v_\Phi &  M_{H \Delta} v_\Delta
	+\frac{1}{2}\lambda^\prime_{H\Phi} v_\Phi^2
	&  
	\frac{1}{2} M_{H \Delta} v\\
	- \frac{1}{2} M_{\Phi \Delta} v_\Phi & \frac{1}{2} M_{H \Delta} v & 
\frac{1}{4 v_\Delta} \left( M_{H\Delta} v^2 + M_{\Phi \Delta} v_\Phi^2 \right)\end{pmatrix} .
\label{eq:Z2oddmassmatrix}
\end{align}
This matrix has zero determinant, which means that at least one of the mass
eigenstates is massless.  Despite complex fields, the mass matrix in Eq.~\eqref{eq:Z2oddmassmatrix} is real and symmetric, 
we can rotate this matrix into its diagonal form through a similarity transformation with the orthogonal matrix
$\mathcal{O}^D$,
\begin{equation}
({\cal O}^D)^{\rm T}\cdot {\mathcal M}_0^{\prime 2} \cdot {\cal O}^D =
{\rm Diag}(0, m^2_D, m^2_{\widetilde\Delta}) \; .
\label{eq:OTM0sqO2}
\end{equation}
The relation between interaction and mass states is given
by $\mathcal{G} = {\cal O}^D\cdot \{ \widetilde{G}^p, D, \widetilde{\Delta}
\}^{\rm T}$. 
The first zero eigenvalue in Eq.~\eqref{eq:OTM0sqO2} corresponds to $\widetilde{G}^{p,m}$, 
the would-be Goldstone boson to be absorbed by $W^{\prime \, (p,m)}$, the complex gauge bosons of $SU(2)_H$.
Here we assume the hierarchy $m^2_D < m^2_{\widetilde{\Delta}}$. Note that we
strictly avoid degenerate masses to simplify the analysis when $D$ is the dark
matter candidate. However, from the mass expressions given below, one can see
that very specific parameter choices are necessary to make the two massive
states degenerate. The masses of the two physical massive eigenstates are given by
\begin{equation}
\label{darkmattermass}
M^2_{D, {\widetilde \Delta}} = \frac{-B \mp \sqrt{B^2 - 4 A C}}{2A} \; , 
\end{equation}
where
\begin{align}
\label{ABC}
A &= 8 v_\Delta \; , \nonumber \\ 
B & = - 2 \left[ M_{H\Delta} \left( v^2 + 4 v_\Delta^2 \right) + M_{\Phi \Delta}
	\left( 4 v_\Delta^2 + v_\Phi^2 \right) + 2 \lambda^\prime_{H\Phi} v_\Delta
\left( v^2 + v_\Phi^2 \right) \right] \;, \\
C & = \left( v^2 + v_\Phi^2 + 4
	v_\Delta^2 \right) \left[ M_{H \Delta} \left( \lambda^\prime_{H\Phi} v^2 + 2
	M_{\Phi \Delta} v_\Delta \right) + \lambda^\prime_{H\Phi} M_{\Phi \Delta}
v_\Phi^2  \right] \; .\nonumber
\end{align}

The lightest state between $H^\pm$ and $D$, if lighter than every other
$\mathcal{Z}_2$-\emph{odd} states, has the possibility to become the DM candidate.
However, an electrically charged DM candidate such as $H^\pm$ is undesirable.
For this reason, we will concentrate on parameter space where $m_{H^\pm} >
m_D$.

\subsubsection*{Gauge Bosons}

After SSB, the gauge bosons that acquire mass terms are the $B$, $X$, and all
the components of $W$ and $W^\prime$. The charged $W^\pm$ gauge bosons remains
completely SM-like with its mass given by $M_W = gv/2$. The $W^{\prime \, p} = ( W^{\prime \, m} )^*$
does not mix with the SM $W^\pm$ and acquires a mass
given by
\begin{equation}
m^2_{W^{\prime (p,m)}}  = \frac{1}{4} g^2_H \lee v^2 + v^2_\Phi + 4 v^2_\De \rii.  \; 
\label{eq:Wppmmass}
\end{equation}
The remaining gauge bosons, $B$, $W^3$, $W^{\prime 3}$ and $X$ have mixing terms. We
can write their mass terms as a 4$\times$4 matrix using the basis 
${\mathcal V}^\prime = \{B,W^{3},W^{\prime 3},X\}^{\rm T}$,
\begin{equation}
\label{M1sq2}
{\mathcal M}_1^2 =     \begin{pmatrix}
	\frac{g^{\prime 2} v^2 }{4} + M_Y^2 & - \frac{g^{\prime} g \, v^2 }{4}  &  \frac{g^{\prime} g_H v^2 }{4} & \frac{g^\prime g_X v^2}{2} + M_X M_Y \\
	- \frac{g^{\prime} g \, v^2 }{4} & \frac{ g^2 v^2 }{4} & - \frac{g g_H v^2 }{4} 
	& - \frac{ g g_X v^2  }{2} \\
	\frac{g^{\prime} g_H v^2 }{4} & - \frac{g g_H v^2 }{4}   & \frac{g^2_H  \lee v^2 + v^2_\Phi \rii }{4}  & 
	\frac{g_H g_X \lee v^2 - v^2_\Phi \rii }{2} \\ 
	\frac{g^\prime g_X v^2}{2} + M_X M_Y & - \frac{ g g_X v^2  }{2}
	& \frac{g_H g_X \lee v^2 - v^2_\Phi \rii }{2} & g_X^2 \left( v^2 + v^2_\Phi \right) + M_X^2
\end{pmatrix} \; ,
\end{equation}
where $M_X$ and $M_Y$ are the two Stueckelberg mass parameters~\cite{Stueckelberg:1938zz,Ruegg:2003ps,Kors:2005uz,Kors:2004iz,Kors:2004ri, Kors:2004dx,Feldman:2007nf,Feldman:2007wj,Feldman:2006wb} 
introduced for $U(1)_X$ and $U(1)_Y$ respectively.
This mass matrix has zero-determinant, meaning that there is at least one
massless state that can be identified with the photon.  The remaining three
states are massive in general. One of them, the $Z$, is related to the SM
gauge boson $Z^{\rm SM}$, and the other two are the extra gauge bosons
$Z^\prime$ and $Z^{\prime\prime}$. As in the neutral scalars case, we
can diagonalize this mass matrix by an orthogonal rotation matrix
$\mathcal{O}^{G}_{4\times 4}$ such that 
${\mathcal V}^\prime = \mathcal{O}^{G}_{4\times 4}\cdot\{A,Z,Z^\prime,Z^{\prime\prime}\}^{\rm T}$. 
We will also use $Z_i$ with $i=1,2,3$ for $Z$, $Z^\prime$, $Z^{\prime\prime}$ 
respectively in the following.
As noted in Ref.~\cite{Huang:2019obt}, one can justify the parameter choice $M_Y = 0$ 
by considering the electric charges of the fermions. Otherwise, the
neutrinos would not be neutral and all the SM electric charges would receive a
correction that grows with $M_Y$. Therefore, hereafter we will consider $M_Y =
0$. This choice makes possible to rotate the first and second rows and columns of Eq.~\eqref{M1sq2} (with $M_Y$ set to 0)
using the Weinberg angle in a SM-like manner. Applying the $4\times 4$ rotation
\begin{equation}
\label{eq:rot4by4dec}
{\cal O}^{W} =
\begin{pmatrix}
    c_W & -s_W & 0 & 0 \\
    s_W & c_W & 0 & 0 \\
    0 & 0 & 1 & 0 \\
    0 & 0 & 0 & 1
\end{pmatrix}
\end{equation}
as $({\cal O}^{W})^{\rm T}\cdot \mathcal{M}_1^2 (M_Y=0)\cdot{\cal O}^{W} $, we find the mass matrix 
\begin{align}
{\cal M}^2_Z = 
\begin{pmatrix}
    0 & 0 & 0 & 0\\
    0 &
    M_{Z^\text{SM}}^2 &
        - \frac{g_H v }{2} M_{Z^\text{SM}} &
        - g_X v M_{Z^\text{SM}} \\
    0 &
        - \frac{g_H v}{2} M_{Z^\text{SM}} &
        \frac{g_H^{2} \left(v^{2} + v_\Phi^{2}\right)}{4} &
        \frac{g_X g_H \left(v^{2} - v_\Phi^{2}\right)}{2}\\
    0 &
        - g_X v M_{Z^\text{SM}} &
        \frac{g_X g_H \left(v^{2} - v_\Phi^{2}\right)}{2} &
        g_X^{2} (v^{2} + v_\Phi^{2}) + M_X^{2}
\end{pmatrix} \, ,
\label{eq:MgaugeSMrot}
\end{align}
where $M_{Z^{\rm SM}} = \sqrt{g^2 + g^{\prime 2}} v/2$ is the SM gauge boson 
$Z^{\rm SM}$ mass. Given the form
of the rotation matrix ${\cal O}^{W}$ we can identify the first component of
the basis of this matrix with the photon, which is immediately massless, and
the second with the $Z^{\rm SM}$. The new intermediate basis in this case is
$\mathcal{V}'_Z=\{A,Z^{SM}, W^{\prime 3}, X\}^{\rm T}$. We can rewrite the
original rotation matrix as the product of two matrices $\mathcal{O}^{G}_{4
\times 4}(M_Y=0)
= \mathcal{O}^W \cdot \mathcal{O}^Z$, where the matrix $\mathcal{O}^Z$ diagonalizes
$\mathcal{M}^2_Z$ in Eq.~\eqref{eq:MgaugeSMrot}. 
In that case we can relate the mass eigenstates with the
intermediate states as $\mathcal{V}'_Z = \mathcal{O}^Z\cdot \{A, Z, Z',
Z''\}^{\rm T}$.
Hereafter, we will call $\mathcal{O}^G$ to the non-diagonal $3\times 3$ part of $\mathcal{O}^Z$,
such that $\mathcal{O}^Z_{j+1,k+1} = \mathcal{O}^G_{j,k}$ with $j$ and $k=1,2,3$, as explicitly
shown in Eq.~(6) of Ref.~\cite{Huang:2019obt}.
Note that the photon $A$ remains the same between
the intermediate states $\mathcal{V}'_Z$ and the mass eigenstates. This
necessarily means that the only non-zero element in the first column and row
of $\mathcal{O}^Z$ is $\mathcal{O}^Z_{1,1} = 1$.

Interestingly, the only gauge boson that acquires mass contributions from the
three non-zero VEVs is the $W^{\prime (p,m)}$ with its mass given in
Eq.~\eqref{eq:Wppmmass}.

\section{Accidental Discrete Symmetry ($h$-parity) and Dark Matter Candidate}
\label{sec:HiddenP}

As mentioned in the previous session, the stability of the scalar dark matter candidate 
in this model is protected by the accidental discrete $\mathcal Z_2$ symmetry in the scalar potential which 
is automatically implied by the $SU(2)_L \times U(1)_Y \times SU(2)_{H} \times U(1)_X$ gauge symmetry. 
Due to its special vacuum alignment where the $H_{2}$ field does not acquire a VEV, 
the accidental $\mathcal Z_2$ symmetry remains intact after SSB. 
It was argued in~\cite{Huang:2015wts} that there is no gauge invariant higher dimensional operator
that one can write down which can lead to the decay of DM candidate in G2HDM. The presence of the
accidental discrete $\mathcal Z_2$ symmetry after SSB reinforces such argument.

This discrete $\mathcal Z_2$ symmetry in G2HDM that we observe here 
is kind of peculiar in the sense
that different components of the $SU(2)_H$ doublets $H$ and $\Phi_H$, and triplet $\De_H$ have opposite parity.  Thus for dark matter physics it is mandatory to give VEVs to those scalars with even parity. 
Otherwise the $\mathcal Z_2$ symmetry will be broken spontaneously which will lead to 
no stable DM as well as the domain wall problem in the early universe. 
Another peculiar feature of this ${\mathcal Z}_2$ symmetry is that it acts on the complex fields.
We will refer this accidental discrete $\mathcal Z_2$ symmetry 
as the hidden parity ($h$-parity) in G2HDM in what follows. 

This $h$-parity can actually be extended to the whole renormalizable Lagrangian of G2HDM, including the gauge, scalar and Yukawa interactions. For example, 
while the SM $W^\pm$ and all the neutral gauge 
bosons $\gamma$, $Z_i$ are always coupled to a pair of SM fermions 
$\bar f f^{(\prime)}$ or a pair of new heavy fermions $\bar f^H f^{(\prime) H}$, 
the $W^{\prime (p,m)}$ always couples to one SM fermion and one new heavy fermion 
$\bar f^H f^\prime$ or $\bar f f^{\prime H}$. 
Similar features can be observed in the gauge-Higgs sector and the Yukawa couplings in G2HDM. For instance, 
while $\gamma$, $Z_i$ and $W^\pm$ are always 
coupled to a pair of $\mathcal Z_2$-even scalars or a pair of $\mathcal Z_2$-odd scalars,
the $W^{\prime (p,m)}$ always couples to one $\mathcal Z_2$-even and one $\mathcal Z_2$-odd scalars.
Also, $h_i$ always couple to either $\bar f f$ or $\bar f^H f^H$, while
for the new Yukawa couplings, the dark matter $D$ $(D^*)$ always couples
with $\bar u u^H$ and $\bar d^H d$ ($\bar u^H u$ and $\bar d d^H$)
and the charged Higgs $H^+$  $(H^-)$ always couples with $\bar u d^H$ and $\bar u^H d$ 
($\bar d^H u$ and $\bar d u^H$), {\it etc}. 
Therefore, besides the $\mathcal Z_2$-even/odd scalars discussed in the previous section, 
one is naturally lead to assign $W^{\prime (p,m)}$ and 
all new heavy fermions $f^H$ to have odd $h$-parity, and 
all SM gauge particles including the additional neutral gauge bosons to have even $h$-parity.
A summary of the $h$-parity for all the fields in G2HDM is collected in Table~\ref{tab:Z2Eff}.
\begin{table}[htbp!]
	\begin{tabular}{|c|c|c|}
		\hline
		Fields & $h$-parity\\
		\hline
		 $h$, $G^{\pm,0}$, $\phi_{2}$, $G^0_H$, $\delta_{3}$, $f$, $W^{\mu}_{1,2,3}$, 
		 $B_{\mu}$, $X^{\mu}$, $W^{\mu\prime}_{3}$, $G^{\mu a}$   & 1 \\
		 \hline
		 $G^{p,m}_H$, $H_2^0$, $H_2^{0 *}$, $H^{\pm}$, $\Delta_{p,m}$, $f^{H}$, 
		 $W^{\mu \prime}_{1,2}$ & $-1$ \\
		\hline 
	\end{tabular}
	\caption{Classification of all the fields in G2HDM under $h$-parity.} 
	\label{tab:Z2Eff}
\end{table}

Thus besides the two well-known accidental global symmetries of baryon number and lepton number inherited from the SM,
there is also an accidental discrete $\mathcal Z_2$ symmetry in G2HDM. 
Other than protecting the stability of the lightest electrically neutral $\mathcal Z_2$-odd particle 
to give rise a DM candidate, 
this accidental $\mathcal Z_2$ symmetry also provides natural flavor conservation laws for 
neutral currents~\cite{Glashow:1976nt,Paschos:1976ay} 
at the tree level for the SM sector in G2HDM~\cite{Huang:2015wts},
as described in previous paragraph.
While it is important to unravel if the $h$-parity in G2HDM has a deeper origin from 
a larger theoretical structure, for example  like grand unification or supersymmetry or 
braneworld, we will not pursue further here.


In principle, any electrically neutral $\mathcal{Z}_2$-odd neutral particle can be a DM candidate 
({\it e.g.} 
the heavy neutrinos $\nu^H$, the
complex scalar mass eigenstate $D$ and the gauge boson $W^{\prime (p,m)}$).
In this work, we focus on the lightest
${\cal Z}_2$-odd complex scalar field $D$. From Eq.~\eqref{eq:Z2oddmassmatrix}
we know that $D$ is a linear combination of the interaction states $G^p_{H}$,
$H^{0*}_{2}$, and $\Delta_{p}$. Using $\mathcal{G} = {\cal O}^D\cdot \{
\widetilde{G}^p, D, \widetilde{\Delta} \}^{\rm T}$ we can write this linear
combination as
\begin{equation}
\label{eq:Dcomposition}
D = \mathcal{O}^D_{12} G^p_H + \mathcal{O}^D_{22} H^{0*}_2+ \mathcal{O}^D_{32}
\Delta_p \;,
\end{equation}
where $\mathcal{O}^D_{ij}$ represents the $(i,j)$ element of the orthogonal
matrix $\mathcal{O}^D$. The actual values of the elements of this matrix
depend on the actual numerical values of the parameters in
Eq.~\eqref{eq:Z2oddmassmatrix}.

Since a particular dominant component cannot be inferred from
Eq.~\eqref{eq:Z2oddmassmatrix} together with the constraints presented in
Sec.~\ref{subsec:pottheorcons}, we take the approach of considering three
different main compositions. 
Using the rotation matrix elements we can define the $G^p_H$,
$H^{0*}_{2}$ and $\Delta_{p}$ compositions as
$f_{G^p}=(\mathcal{O}^D_{12})^2$, $f_{H_2}=(\mathcal{O}^D_{22})^2$ and
$f_{\Delta_{p}}=(\mathcal{O}^D_{32})^2$ respectively, 
satisfying $f_{G^p}+f_{H_2}+f_{\Delta_p}=1$. 
Our results will be classified in three different cases:
\begin{enumerate}
	\item Inert doublet-like DM for $f_{H_2} > 2/3$,
	\item $SU(2)_H$ triplet-like DM for $f_{\Delta_p} > 2/3$,
	\item $SU(2)_H$ Goldstone boson-like DM for $f_{G^p} > 2/3$.
\end{enumerate}
To avoid cluttering in the following, we will use the more concise terms doublet-like, triplet-like 
and Goldstone-like DM to refer to the above cases of 1, 2 and 3 respectively.

\begin{figure}[htb]
\includegraphics[width=0.5\textwidth]{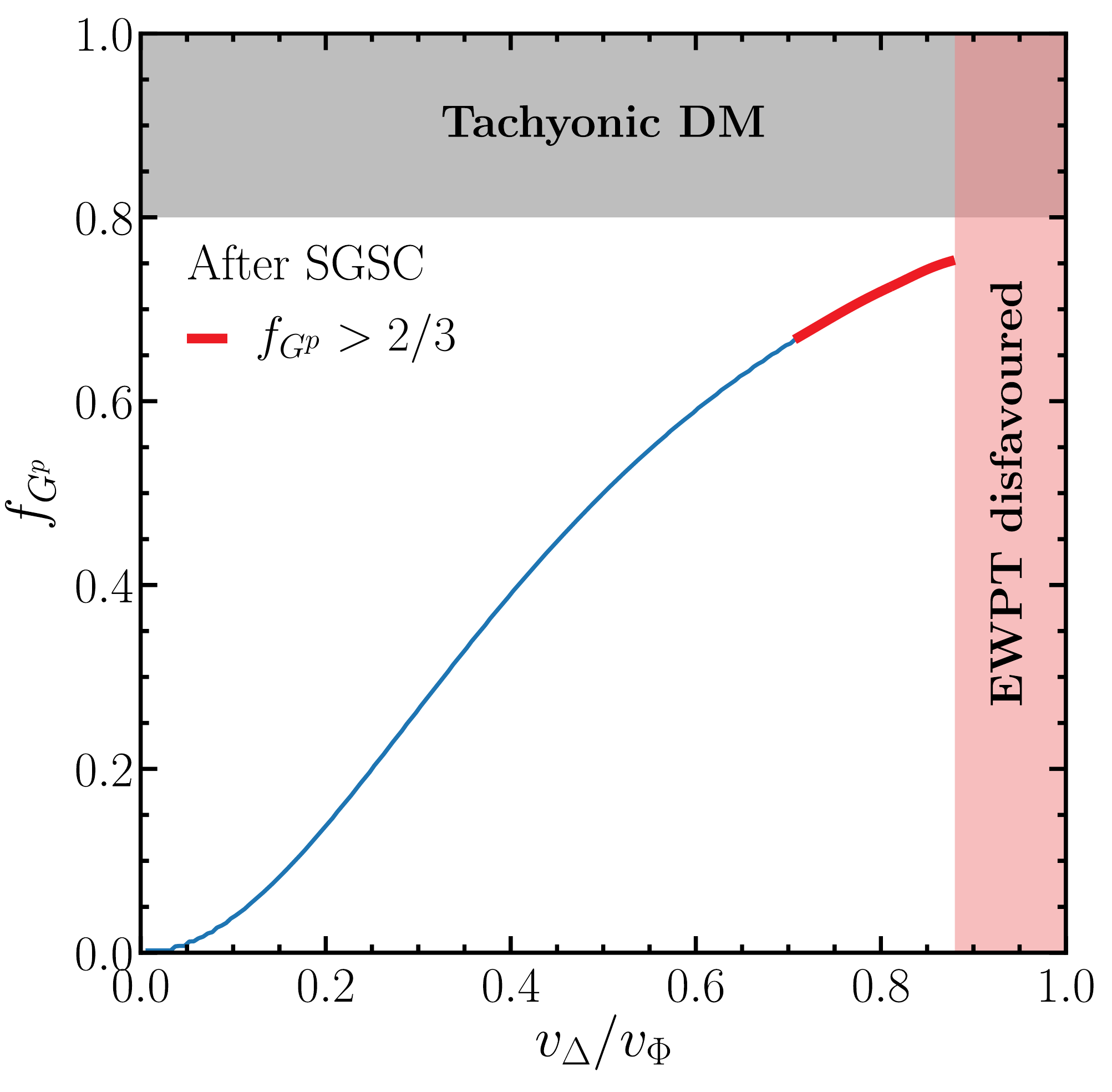}
\caption{\label{fig:vD_vP_fgp}Correlation between the ratio $v_\Delta/v_\Phi$ and 
the composition mixing parameter $f_{G^p}$ for all the DM types after applying constraints 
from the scalar and gauge sectors. 
}
\end{figure}

In order to realize any one of the three cases of the DM discussed above, one needs 
to have its diagonal element in the mass matrix given by
Eq.~\eqref{eq:Z2oddmassmatrix} to be the lightest, while its mixings with the other two 
off-diagonal elements are small. However, the mixing among the other two can be arbitrary.
Take the Goldstone-like DM as an example.
It is easy to note that the (1,1) and (3,3) elements of the mass matrix in
Eq.~\eqref{eq:Z2oddmassmatrix} have a see-saw behaviour controlled by the
value of $v_\Delta$. 
The (2,2) element remains almost unaffected thanks to the
term proportional to large $v_\Phi^2$.  Goldstone-like DM is characterized by
a large value in the (1,1) element of Eq.~\eqref{eq:Z2oddmassmatrix} when
compared to the (1,2) and (1,3) elements, given by $\lambda^\prime_{H\Phi}v v_\Phi /2$ and
$-M_{\Phi\Delta}v_\Phi/2$ respectively, so as to suppress the mixing effects. The size of the
(1,2) element is not relevant since it is proportional to the smaller term $v v_\Phi$ as compared with
both the (1,1) and (2,2) elements which are always much larger. The
difference in size between the (1,1) and (1,3) elements is best measured by taking 
the ratio between them which is roughly about $2v_\Delta/v_\Phi$. In
other words, the $v_\Delta/v_\Phi$ ratio controls the Goldstone boson
composition of the DM mass eigenstate. This is illustrated in
Fig.~\ref{fig:vD_vP_fgp}, where the correlation between the ratio 
$v_\Delta/v_\Phi$ and the composition mixing parameter $f_{G^p}$ is  shown for all DM types.
The small arc in the correlation curve with $f_{G^p} > 2/3$ is highlighted by red color
indicating only a small parameter space is allowed for Goldstone-like DM.
Note that when the ratio $v_\Delta/v_\Phi$ grows close to 1, 
EWPT disfavors the presence of relatively light $Z^\prime$ state with larger mixing with the SM $Z$.
Values of $f_{G^p}$ larger than $\sim 0.8$ are accessible only through
negative $M_{\Phi\Delta}$ resulting in tachyonic DM mass.
Therefore to realize a Goldstone-like DM with $f_{G^p} > 2/3$ 
in the following numerical analysis, 
one needs to do some fine-tunings in the parameter space.
For the inert doublet-like and triplet-like DM, no fine-tunings are required.

\section{Dark Matter Experimental Constraints}
\label{sec:constraints}

To determine WIMP DM properties, one can measure the DM-SM interactions via
several approaches.  Conventionally, DM direct detection, DM indirect
detection, and collider search are used for the hunting for DM.  Thus far 
null signals were reported from all these experimental efforts 
and only allowed regions have been shown in the DM model parameter space.
On the other hand, the DM relic density
measurement can indeed give a signal region which can constrain the parameter space 
of DM model in a significant way.  Hence, our strategy is to determine the parameter space in
G2HDM allowed by the current relic density measurement, and the limits deduced 
from DM direct detection, DM indirect detection, and collider search (at the LHC).  
In this section we briefly describe each of these experimental constraints used in this analysis. 

\subsection{Relic Density}
\label{sec:relic}

It is fascinating to wonder about the thermal history of DM based on all our
current knowledge of physics.  The simplest scenario is that a WIMP maintains
its thermal equilibrium with the SM sector before freeze-out and 
the DM number density can be described by a Boltzmann distribution.  
Therefore, the DM mass determines its abundance before freeze-out.  
As in most WIMP theories, owing to the small DM-SM couplings, 
the relic density comes out too large and the correct abundance can be only achieved by
some specific mechanisms. The mechanisms to reduce the thermal DM relic
density in the G2HDM can be both from DM annihilation and
also from coannihilation with heavier ${\cal Z}_2$-odd particles.
Coannihilation only
happens if the next lightest ${\cal Z}_2$-odd particles are slightly
heavier than DM (usually $\lesssim 10\%$) so that its number density at the
temperature higher than freeze-out does not suffer a large Boltzmann
suppression.  In our setup, the heavier ${\cal Z}_2$-odd scalar
$\widetilde{\Delta}$, the charged Higgs, new heavy fermions, or
gauge boson $W^{\prime (p,m)}$ can coannihilate with the DM candidate $D$.
Additionally, the SM Higgs and $Z$ resonance can play an important role
for the doublet-like DM while there is no $Z$ resonance in the
triplet-like and Goldstone-like DM cases because both
$\Delta_H$ and $\Phi_H$ are SM singlets. As we will
see later, the couplings between DM and some of the mediators in G2HDM could
be suppressed by mixings or cancellations.

 The scalar $\widetilde{\Delta}$ and the DM candidate
$D$ come from the same mass matrix. The splitting between their masses is
mostly controlled by the second term in the numerator of Eq.~\eqref{darkmattermass}.
Hence, coannihilation between DM and $\widetilde{\Delta}$
can only happen if $M_{\Phi\Delta}\gtrsim\mathcal{O}(10\gev)$ and
$v_\Phi\gtrsim 70\tev$. However, this condition also makes DM masses at around 
$\mathcal{O}(1\tev)$ or larger. 

For the doublet-like DM case, the mass of the DM candidate is close to
the mass of the charged Higgs with the splitting approximately given by
\begin{equation}
\label{eq:mcHsplit}
m^{2}_{H^{\pm}} - m^{2}_D \approx - \frac{1}{2} \lambda'_{H} v^{2} \, ,
\end{equation}
in the approximation where DM mass is dominated by the (2,2) element of the
matrix in Eq.~\eqref{eq:Z2oddmassmatrix}.  For triplet-like DM, the mass
differences between $D$ and the other heavy $\mathcal{Z}_2$-odd scalars are
usually large enough so that coannihilation can be avoided.  
Coannihilation between $D$ and $W^\prime$ occurs for 
DM mass closing to $W^\prime$ mass which is heavy due to large $v_\Phi$.  
As for the resonance, only $h_{1}$ and $h_{2}$ resonances are present. Due to
$\Delta_H$ being an SM singlet,  any $D$ annihilation through $Z$-boson like 
mediator is suppressed by mixings.

To compare against experimental data, we will consider the latest result
from the PLANCK collaboration~\cite{Aghanim:2018eyx} for the relic density,
$\Omega h^2 = 0.120\pm 0.001$. In particular, we will require the parameter space 
of G2HDM to reproduce this well measured value with a $2\sigma$ significance.

\subsection{Direct Detection}

The most recent constraint for DM direct search is given by
the XENON1T collaboration~\cite{Aprile:2018dbl}.  The null signal 
result from this search puts the most stringent limit on DM nucleon cross section so far, 
especially for the DM mass that
lies between 10~GeV to 100~GeV.  The XENON1T collaboration excluded DM-nucleon
elastic cross sections above $10^{-46} \text{~cm}^{2}$ for a DM particle with
mass around 25~GeV.

In models with isospin violation (ISV), DM interactions with proton
and neutron can be different and the ratio between the DM-neutron and DM-proton
effective couplings, $f_n/f_p$, can have values that
differ from 1 significantly depending on the model parameters. 
In particular, for a target made of xenon, the ratio $f_n/f_p\approx-0.7$ corresponds
to maximal cancellation between proton and neutron
contributions~\cite{Feng:2011vu}.

For instance, if DM interacts with nucleons mediated by the $Z$ boson, the
strength is characterized by the electric charge and the third generator
$T_{3}$ of $SU(2)_{L}$ group. The vectorial coupling of quark $q$ ($u$ or $d$-type) 
to the SM $Z$ boson in G2HDM is 
\bea
\label{eq:qqsz}
g^{V}_{\bar{q} q Z} = \frac{i}{2} \left[ \frac{g}{c_{W}} \left(T_{3} - 2 Q_q s^{2}_{W} \right) \mathcal{O}^{G}_{11} + g_{H} T^{\prime}_{3} \mathcal{O}^{G}_{21} + g_{X} X \mathcal{O}^{G}_{31} \right] \, .
\eea
Due to different $Q_q$, $T_3$, $T'_3$ and $X$ charges, this coupling is expected to
vary depending on the quark $q$ being $u$ or $d$-type.
Hence, the $Z$ boson interacts with proton and neutron differently 
and the $f_n/f_p$ can be different from 1.
For the case of DM with a non-negligible doublet composition   
we found that DM can couple to proton or neutron differently 
via $Z_i$ boson exchange and it leads to ISV.

Generally speaking, exact cancellation between neutrons and protons 
is expected to be in a tiny region of parameter space.  Nevertheless,
the G2HDM doublet-like DM can have a much wider distribution of $f_n/f_p$.  
Furthermore, a more subtle interference between the contributions from 
neutral gauge and Higgs boson exchange can result in 
two different scattering cross sections for DM ($D$) and antiDM ($D^*$) with
neutrons.  Such a difference may compensate for any cancellation caused by
ISV.  While one can expect a negligible cancellation 
from ISV effects
in most of the G2HDM parameter space, we consider these effects in all the DM
scenarios.
To include ISV effects, one has to compute the DM-nucleus elastic scattering
cross section $\sigma_{D \mathcal N}$
%
\begin{equation}
\sigma_{D\mathcal{N}}=\frac{4\mu^2_{\mathcal{A}}}{\pi}
\left[ 
f_p \mathcal{Z} + f_n (\mathcal{A} - \mathcal{Z})
\right]^2 \, ,
\label{eq:sigmaTH}
\end{equation}
where $\mathcal N$ stands for a nucleus with mass number $\mathcal{A}$ and proton number $\mathcal{Z}$. 
For definiteness, we will ignore all the isotopes of xenon and fix $\mathcal A$ and $\mathcal{Z}$
to 131 and 54 respectively in this work.
We obtain the effective couplings $f_p$ and $f_n$ by using
\texttt{micrOMEGAs}~\cite{Belanger:2018mqt}.  The DM-nucleon reduced mass
is denoted as $\mu_{\mathcal{A}}= m_D m_\mathcal{A}/(m_D +m_\mathcal{A})$.
On the other hand, the limit published by XENON1T is for the nucleon with isospin
conserving assumption $f_n=f_p$.
To reconstruct the XENON1T results at the nucleus level for general value of the ratio $f_n/f_p$,
we use the following expression
\begin{equation}
\sigma^{\rm X1T}_{D\mathcal{N}}=\sigsip({\rm X1T})\times  
\frac{\mu^2_{\mathcal{A}}}{\mu^2_{p}} \times \left[ \mathcal{Z} + \frac{f_n}{f_p} 
\left({\mathcal{A} - \mathcal{Z}} \right) \right]^2 \, ,
\label{eq:sigmaEXP}
\end{equation} 
where $\mu^2_{p}$ is the DM-proton reduced mass. 
In this work, we use Eq.~\eqref{eq:sigmaEXP} to constrain our direct detection
prediction.

\begin{figure}[!phtb]
  \begin{minipage}{0.46\textwidth}
      \includegraphics[page=13]{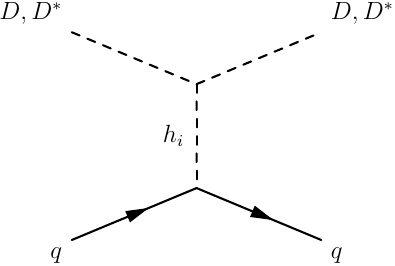}
  \end{minipage}
  \begin{minipage}{0.46\textwidth}  
      \includegraphics[page=14]{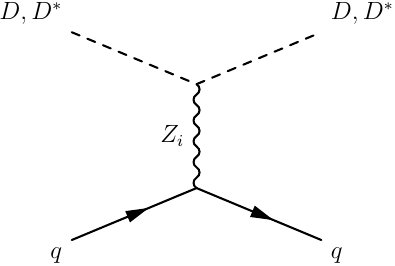}
  \end{minipage}
\caption{The dominant Feynman diagrams with the $\mathcal Z_2$-even Higgs bosons (left) 
and neutral gauge bosons (right) exchange for direct detection of DM. 
}
\label{fig:DDFeynman}
\end{figure}

Since we are dealing with complex scalar DM, we need to consider the
antiDM interaction with the nucleon. The DM-nucleon interaction and antiDM-nucleon interaction in general can be quite different. When the mediators are heavy enough, one can integrate them out to obtain 
effective interactions for the DM and nucleon. 
The spin independent interaction for complex scalar DM can be written in terms of effective operator as~\cite{Belanger:2008sj}
\begin{eqnarray}
\label{eq:scalar_si_lgrgn}
  \mathcal{L}_{D} &=& 2 \lambda_{N,e} M_D D D^*  \bar{\psi}_N\psi_N
  + i \lambda_{N,o} \left( D^* \overset{\longleftrightarrow}{\partial_\mu} D \right)
\bar{\psi}_N\gamma^\mu \psi_N
\end{eqnarray}
where the $\psi_N$, $\lambda_{N,e}$, and $\lambda_{N,o}$ denote the nucleon
field operator, the coupling of even operator, and the coupling of odd
operator respectively. The  effective coupling of DM (antiDM) with the nucleon is given by 
\begin{eqnarray}
  \lambda_N =\frac{\lambda_{N,e} \pm
  \lambda_{N,o}}{2} \; ,
  \label{eq:odd-even}
\end{eqnarray}
where the plus (minus) sign stands for DM-nucleon (antiDM-nucleon) interaction. The first term in the right hand side of
Eq.~\eqref{eq:scalar_si_lgrgn} represents the \emph{even} operator interaction
between DM and the nucleon. It is called \emph{even} operator because when
one exchanges $D$ with $D^*$, the
interaction stays the same. On the other hand, under a similar exchange
between $D$ and $D^*$ the second term flips sign. Thus, it is called
\emph{odd}
operator. As a result, the interaction strength between DM-nucleon
and antiDM-nucleon will not be the same and it is given by
Eq.~\eqref{eq:odd-even}. 
Hence the numerical value of $\sigma_{D^* \cal N}$ is in general not equal to 
$\sigma_{D \cal N}$ given by Eq.~\eqref{eq:sigmaTH} because the effective
couplings $f_p$ and $f_n$ for $D$ are not the same as those for $D^*$.

The Feynman diagrams of the dominant contribution to describe DM-quark
interactions in the G2HDM are shown in Fig.~\ref{fig:DDFeynman}.  The left
panel is the $t$-channel with three Higgs bosons exchange while the right panel
is the $t$-channel with three neutral gauge bosons exchange. Thus G2HDM captures
the features from both the Higgs-portal and vector-portal DM models in the literature.

Note that the doublet-like DM in this model has a large scattering cross section  
because the vertex $DD^*Z$ is governed by the SM coupling as shown in 
Eq.~\eqref{eq:g_d_ddz1},  
and hence \textit{not} suppressed by any mixing angle. 
Due to the additional contributions from other heavy gauge bosons ($Z^\prime$ and
$Z^{\prime\prime}$), 
the DM neutron cross section is three orders of magnitude larger than the DM proton cross section. 
This ISV effect is also observed in triplet-like DM ($\Delta_{p}$) 
but the ISV occurs mildly due to the mixing suppression between the DM ($\Delta_{p}$) and $Z$ coupling. 
This suppression makes the $Z$ exchange contribution comparable with
$Z^\prime$ and $Z^{\prime\prime}$ as well as the contribution from the SM Higgs $h_{1}$ exchange. 

\subsection{Indirect Detection: Gamma-ray from dSphs}

Excluding the early universe, DM at the present may also annihilate to SM
particles significantly at the halo center where DM density is dense enough to
produce cosmic rays or photons which can be distinguished from those standard
astrophysical background.
Such a measurement is known as DM indirect detection.
As long as indirect detection constraints are concerned, the continuum
gamma-ray observations from dwarf spheroidal galaxies (dSphs) can usually
place a robust and severe limit on the DM annihilation cross section for DM
masses larger than 10~GeV~\cite{Fermi-LAT:2016uux}.
This is owing to two advantages of searching DM at the dSphs.
First, the dSphs provides an almost background-free system because they are
faint but widely believed to be DM dominated systems.
Second, their kinematics can be precisely measured, hence the systematical
uncertainties from DM halo can be controlled.
Therefore, in this work we will only use the dSphs constraints implemented in
\texttt{LikeDM}~\cite{Huang:2016pxg} to evaluate the $\chi^2$ statistics of
our model based on Fermi Pass 8 data (photon counts), recorded from
August 4th 2008 to August 4th 2015. 
The two-dimensional $2\sigma$ criteria is taken to be $\Delta\chi^2=5.99$ in
our study.

The standard gamma-ray fluxes produced from DM annihilation at the dSphs halo
is given by 
\begin{equation}
\label{eq:idflux}
\frac{d\Phi_\gamma}{dE_{\gamma}}={\frac{\sigmav}{8{\pi}m_{D}^{2}} \times J\times 
\sum_{\texttt{ch}} {\rm BR}(\texttt{ch})\times \frac{dN^{\texttt{ch}}_{\gamma}}
{dE_{\gamma}}} \, ,
\end{equation}
where $J={\int}dld{\Omega}{\rho}(l)^{2}$ is the so-called $J$-factor, 
which integrates along the line-of-sight $l$ with the telescope opening angle 
given by $\Omega$. The DM density distribution is denoted as $\rho(l)$.
Here, we take 15 dSphs and their $J$-factors as the default implementation in \texttt{LikeDM}.  
The index $\texttt{ch}$ runs over all the DM annihilation channels.  
The annihilation branching ratio ${\rm BR}(\texttt{ch})$ and energy spectra
$dN^{\texttt{ch}}_{\gamma}/dE_{\gamma}$ are computed by using
\texttt{micrOMEGAs} and \texttt{PPPC4}~\cite{cirreli:pppc4}, respectively.  

Similar to Higgs portal models, an inert Higgs DM in our setup can only
annihilate to the SM fermions via Higgs portal or $Z$-boson.  On the other
hand, if the DM is heavier than $m_W$ then the four points interaction
$DD^*W^+W^-$ can have a higher photon flux to be tested.

\subsection{Collider Search}

\subsubsection*{Mono-jet Search}

\begin{figure}
\begin{center}
  \begin{minipage}{0.43\textwidth}
      \includegraphics[page=1]{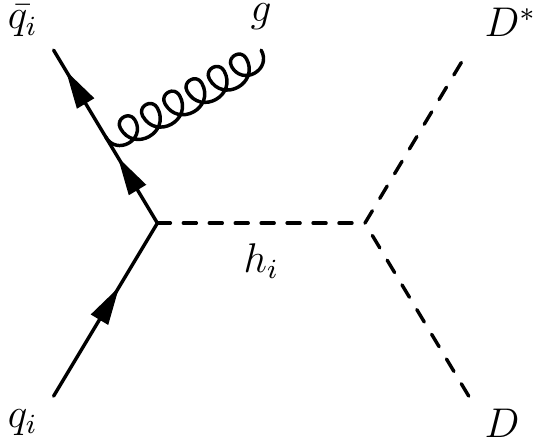}
    \label{mono_hi}
  \end{minipage}
  \begin{minipage}{0.43\textwidth}  
      \includegraphics[page=2]{feynmp_standalone.pdf}
    \label{mono_zi}
  \end{minipage}\\
  \vspace{0.5cm}
  \begin{minipage}{0.43\textwidth}   
      \includegraphics[page=3]{feynmp_standalone.pdf}
\end{minipage}
\begin{minipage}{0.43\textwidth}   
      \includegraphics[page=4]{feynmp_standalone.pdf}
            \label{mono_qh_2}
        \end{minipage}
        \caption      
        {\small The representative Feynman diagrams of leading contributions for mono-jet at the LHC in the G2HDM, where $h_i=h_1, h_2, h_3$ and $Z_i=Z, Z', Z''$.
        } 
        \label{fig:monojet}
\end{center}
\end{figure}

DM particles could be produced copiously  at  colliders.
Unfortunately, DM can not be detected on its own since it would pass through
detectors without leaving any trace. 
Therefore, one should look for the DM production associated with visible SM particles. At the LHC, the signal of an energetic jet from initial state radiation that balances the momentum of undetected DM, usually referred to mono-jet signal, is one of the sensitive channels to the search for DM. 
As shown in Fig.~\ref{fig:monojet}, the DM pairs are mainly produced in the Feynman diagrams with the exchanges of
$\mathcal Z_2$-even Higgs bosons and neutral gauge bosons in the G2HDM.
For numerical study, we take the parameters allowed by EWPT~\cite{Huang:2019obt} and the XENON1T constraints (to be discussed in Sec.~\ref{sec:result}), and find out that the cross sections are far below the current limits set by ATLAS~\cite{Aaboud:2017phn} and CMS~\cite{Sirunyan:2017jix} collaborations at the LHC.   
 Therefore the mono-jet search would not play any significant role in determining the viable parameter space for DM in G2HDM.

\subsubsection*{Invisible Higgs Decay}

The Higgs boson will decay into a pair of DM when the DM is lighter than half
its mass. This decay channel is known as the invisible decay of the Higgs
boson. At tree level in G2HDM, the partial decay width of the Higgs boson to
pair of dark matter, $h_1 \to D D^{*}$, is given by
\begin{equation}
\Gamma(h_1 \to D D^{*}) = \frac{(\mathcal{O}^{D}_{32})^{4}}{16 \pi m_{h_1}}  
\lambda_{DD^{*}h_1}^2
\sqrt{
      1-\frac{4m^{2}_{D}}{m^2_{h_1}} 
     } \, ,
\end{equation}
where the $\lambda_{DD^{*}h_1}$ coupling depends on the composition of the $h_1$. 
For example, for the triplet-like DM case, it can be deduced from 
Eq.~\eqref{eq:g_T_ddh}, \emph{viz.},
\begin{equation}
\lambda_{DD^{*}h_1} \approx
      \mathcal{O}_{11} \lambda_{H\Delta} v +
      \mathcal{O}_{21} \lambda_{\Phi \Delta} v_{\Phi} -
      2 \mathcal{O}_{31} \lambda_{\Delta} v_{\Delta} \, .
\end{equation}

Currently the upper limit on the Higgs invisible decay branching ratio
is rather loose, about $24\%$ at $95\%$ C.L.~\cite{Tanabashi:2018oca} at the LHC. 
Taking $m_D\ll m_{h_1}$ together with SM Higgs total decay width of $13\mev$~\cite{Tanabashi:2018oca}, 
the LHC limit implies an upper bound, 
\begin{equation}
(\mathcal{O}^{D}_{32})^{2}\lambda_{DD^{*}h_1}<5.099\gev \, .
\end{equation}
	However, we found this limit is not as stringent as DM direct detection unless $m_D\lesssim 10\gev$ where DM recoil energy is below the XENON1T threshold.

\section{Numerical Analysis and Results}
\label{sec:result}
\subsection{Methodology}

In order to keep consistency with previous G2HDM studies, in particular the
scalar sector constrains presented in \cite{Arhrib:2018sbz}, we will perform
random scans to generate a sample of points consistent with all the conditions
mentioned there. In our case, we will not keep $v_\Phi$ fixed.
Due to $Z'$ search constraints \cite{Huang:2019obt}, we start our scan range at
$v_\Phi= 20$~TeV.
Considering the energy scale for future colliders, we scan $v_\Phi$
up to 100~TeV~\footnote{For the Goldstone-like DM scenario, the
scan range of $v_\Phi$ is fine-tuned to a smaller range from 20 to 28 TeV in
order to realize this scenario.
}.


We will complete the scan with the free parameters of the gauge sector $g_H$
and $g_X$, while fixing the Stueckelberg mass parameter $M_X = 2$~TeV
corresponds to the heavy $M_X$ scenario discussed in~\cite{Huang:2019obt}.  We
will keep the $g_H$ coupling below 0.1 to avoid the Drell-Yan constraints. The
lower bound of $g_H$ will be decided point by point such that the $W^\prime$
boson is heavier than the DM $D$.
From Eq.~\eqref{eq:Wppmmass}, we can obtain a condition for the minimum value of $g_H$
\begin{equation}
g_{H\text{min}} = \frac{2 m_D}{\sqrt{v^2 + v_\Phi^2 + 4v_\Delta^2}} \, .
\label{eq:ghmincondition}
\end{equation}
Additionally, we will require that the gauge bosons $Z'$ and $Z''$ are both
heavier than the SM-like $Z$ and that the latter has a mass within its
3$\sigma$ measured value of $91.1876\pm0.0021$~GeV.

To keep heavy fermions above detection limits, we will consider their masses to be no less than
1.5~TeV from the searches of SUSY colored particles quoted in the 
PDG \cite{lipi:2017} or $1.2 \times m_D$ from coannihilation consideration.
In addition, we want to keep the new Yukawa couplings,
related to the new heavy fermion masses generically by~\footnote{We note that 
while the Yukawa couplings among the SM fermions and the neutral Higgses
maintain flavor diagonal in G2HDM, the new Yukawa couplings 
are in general not. For simplicity, we have set the unitary mixing matrices 
among different flavors of heavy and SM fermions in the new Yukawa couplings with the 
$\mathcal Z_2$-odd scalars to be the identity matrix.}
$m_{f^H} = y_{f^H} v_\Phi/\sqrt 2$, to be reasonably
small in order to minimize their effects on perturbative unitarity and 
renormalization group running effects. Therefore, we
use the following formula to determine the appropriate Yukawa couplings for
each point in our scan
\begin{equation}
\label{eq:HFYukawa}
y_{f^H} = \text{max}\left[ \frac{1.5\text{ TeV}}{v_\Phi / \sqrt{2}},
\text{min}\left( \frac{1.2\,m_D}{v_\Phi / \sqrt{2}},\; 1 \right) \right]\;.
\end{equation}
Given the size of $v_\Phi$ and the fact that $m_D$ has to be the lightest
$\mathcal{Z}_2$-odd particle, we expect that Eq.~\eqref{eq:HFYukawa} to easily remain
below 1 for all our parameter space.
Thus, in this set up, one expects most coannihilation
contributions are coming from other $\mathcal Z_{2}$-odd particles such as
$\tilde{\Delta}$, $H^{\pm}$ and $W'$.

From these two steps we collect $\sim5$ million points that include numerical
values for model parameters, and results from scalar and gauge bosons masses,
and the elements of three mixing matrices $\mathcal{O}$, $\mathcal{O}^D$ 
and $\mathcal{O}^G$. We pass these numbers to
\texttt{MicrOMEGAs}~\cite{Belanger:2018mqt} to calculate relic density,
DM-nucleon cross section and annihilation cross section at present time.
Finally, the annihilation cross section and annihilation channels composition
are passed to \texttt{LikeDM}~\cite{Huang:2016pxg} for the calculation of
indirect detection likelihood.

Due to the notably less abundant nature of doublet-like solutions compared to
the other two compositions of DM, a scanning dedicated to find doublet-like solutions
was made. For a $M_{H\Delta} \ll v_\Delta$ we can make the (2,2) entry in the mass matrix in 
Eq.~\eqref{eq:Z2oddmassmatrix} smaller than the (3,3) one with the
condition
\begin{equation}
\lambda'_{H\Phi} < \frac{M_{\Phi\Delta}}{2v_\Delta} \, .
\end{equation}
Applying this condition increases the abundance of solutions where the
lightest complex scalar composition is dominated by $H^{0*}_2$. This explains the far
more limited scan range for the parameter $\lambda'_{H\Phi}$ for the
doublet-like DM case. The complete set of parameters scanned and their ranges
can be found in Table~\ref{tab:scanranges}.

Note that in Table~\ref{tab:scanranges}, the different ranges for
$M_{H\Delta}$, $M_{\Phi\Delta}$, $v_\Delta$ and $v_\Phi$ are selected 
for the three cases so that we can easily find the corresponding DM composition.  In
particular, the very different and smaller fine-tuned ranges of $v_\Delta$ and $v_\Phi$ in
the Goldstone-like column are due to this composition being present for
$v_\Delta/v_\Phi\approx 0.8$ but limited by EWPT to be less than $\sim 0.9$, as
demonstrated earlier near the end of Sec.~\ref{sec:HiddenP}.

Before embarking upon the numerical results, we make some comments on the 
Sommerfeld enhancement~\cite{Hisano:2004ds,ArkaniHamed:2008qn} in the DM annihilation 
cross section for indirect detection which may be important whenever
$m_\chi/m_\mathcal{M} > 4 \pi /g^2$. Here
$m_\chi$ and $m_\mathcal{M}$ denote the masses of the fermionic DM $\chi$ and vector mediator
$\mathcal{M}$ respectively 
and $g$ is the gauge coupling. 
In G2HDM,  the DM is a complex scalar $D$ and 
the mediators can be either the Higgses $h_i$ or neutral gauge bosons $Z_i$. 
Since all their masses are quite massive and not too distinct from each other, 
we do not expect significant Sommerfeld enhancement in G2HDM. 
Certainly a more decent study is necessary in order to provide a definite answer. 
Furthermore we will see in our analysis below that the direct detection limit from XENON1T 
will provide more stringent constraints than the current indirect detection results from Fermi-LAT.
We will ignore such effects in the present analysis.

\begin{table}
\begin{tabular}{|c|c|c|c|}
\hline
Parameter & Doublet-like & Triplet-like & Goldstone-like \\
\hline \hline
$\lambda_H$               & [0.12, 2.75]                      & [0.12, 2.75]                      & [0.12, 2.75]      \\
$\lambda_\Phi$            & [$10^{-4}$, 4.25]                 & [$10^{-4}$, 4.25]                 & [$10^{-4}$, 4.25] \\
$\lambda_\Delta$          & [$10^{-4}$, 5.2]                  & [$10^{-4}$, 5.2]                  & [$10^{-4}$, 5.2]  \\
$\lambda_{H \Phi}$        & [$-$6.2, 4.3]                     & [$-$6.2, 4.3]                     & [$-$6.2, 4.3]     \\
$\lambda_{H \Delta}$      & [$-$4.0, 10.5]                    & [$-$4.0, 10.5]                    & [$-$4.0, 10.5]     \\
$\lambda_{\Phi \Delta}$   & [$-$5.5, 15.0]                      & [$-$5.5, 15.0]                  & [$-$5.5, 15.0]  \\
$\lambda_{H \Phi}^\prime$ & [$-$1.0, 18.0]                      & [$-$1.0, 18.0]                    & [$-$1.0, 18.0] \\
$\lambda_H^\prime$        & [$-8\sqrt{2}\pi$, $8\sqrt{2}\pi$] & [$-8\sqrt{2}\pi$, $8\sqrt{2}\pi$] & [$-8\sqrt{2}\pi$, $8\sqrt{2}\pi$] \\
\hline
$g_H$                     & [See text, 0.1]                   & [See text, 0.1]                   & [See text, 0.1]   \\
$g_X$                     & [$10^{-8}$, 1.0]                  & [$10^{-8}$, 1.0]                  & [$10^{-8}$, 1.0]  \\
\hline
$M_{H\Delta}$/GeV         & [0.0, 15000]               & [0.0, 5000.0] & [0.0, 5000.0]      \\
$M_{\Phi \Delta}$/GeV     & [0.0, 5.0]                   & [$-$50.0, 50.0] & [0.0, 700]     \\
$v_{\Delta}$/TeV          & [0.5, 2.0]                       & [0.5, 20.0] & [14.0, 20.0]    \\
$v_{\Phi}$/TeV            & [20, 100]                         & [20, 100] & [20, 28.0]      \\
\hline
\end{tabular}
\caption{Parameter ranges used in the scans mentioned in the
	text. $M_X$ is fixed at 2 TeV in this work and $M_Y$ is set to be zero throughout the scan.
}
\label{tab:scanranges}
\end{table}

\subsection{Results}
\label{analysis}


To ease the discussion of our numerical results, it is useful to divide the DM mass range into 
several regions:
\begin{itemize}
\item[(i)] light DM mass region where annihilation final states of $\ c \bar{c}$ and $\tau^+ \tau^-$ are opened, 
\item[(ii)] the resonance region where DM mass is close to SM $Z$ or Higgs resonance,  
\item[(iii)] the intermediate DM mass from Higgs resonance $m_{h_1}/2$ to $\sim 500\gev$ where DM mainly annihilates 
to $W^+W^-$ and $ZZ$, and
\item[(iv)] the heavy DM mass larger than $500\gev$. 
\end{itemize}

\subsubsection*{Inert Doublet-like DM}
\label{sec:doublet}

\begin{figure}[!htb]
\includegraphics[width=0.48\textwidth]{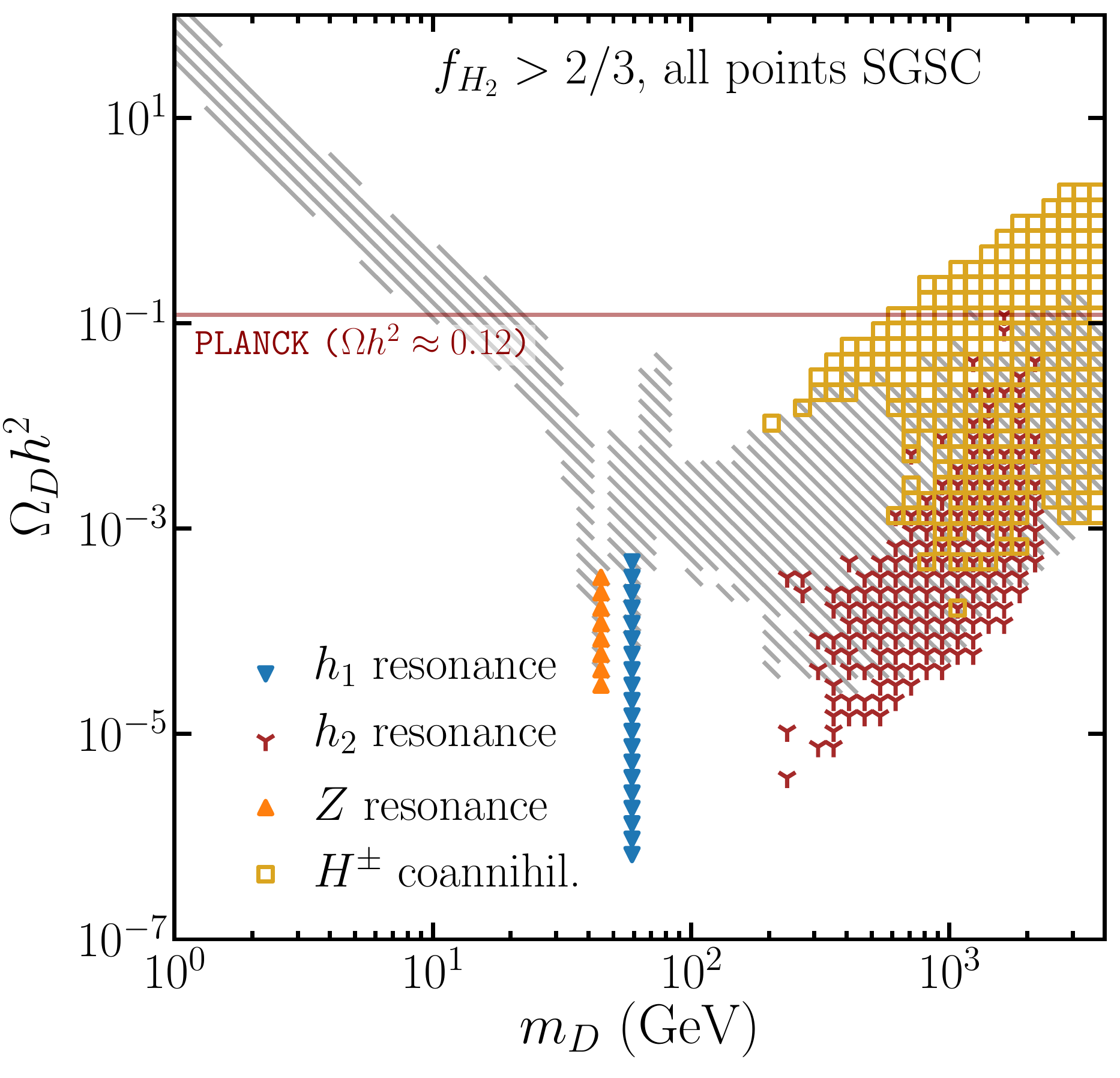}
\includegraphics[width=0.49\textwidth]{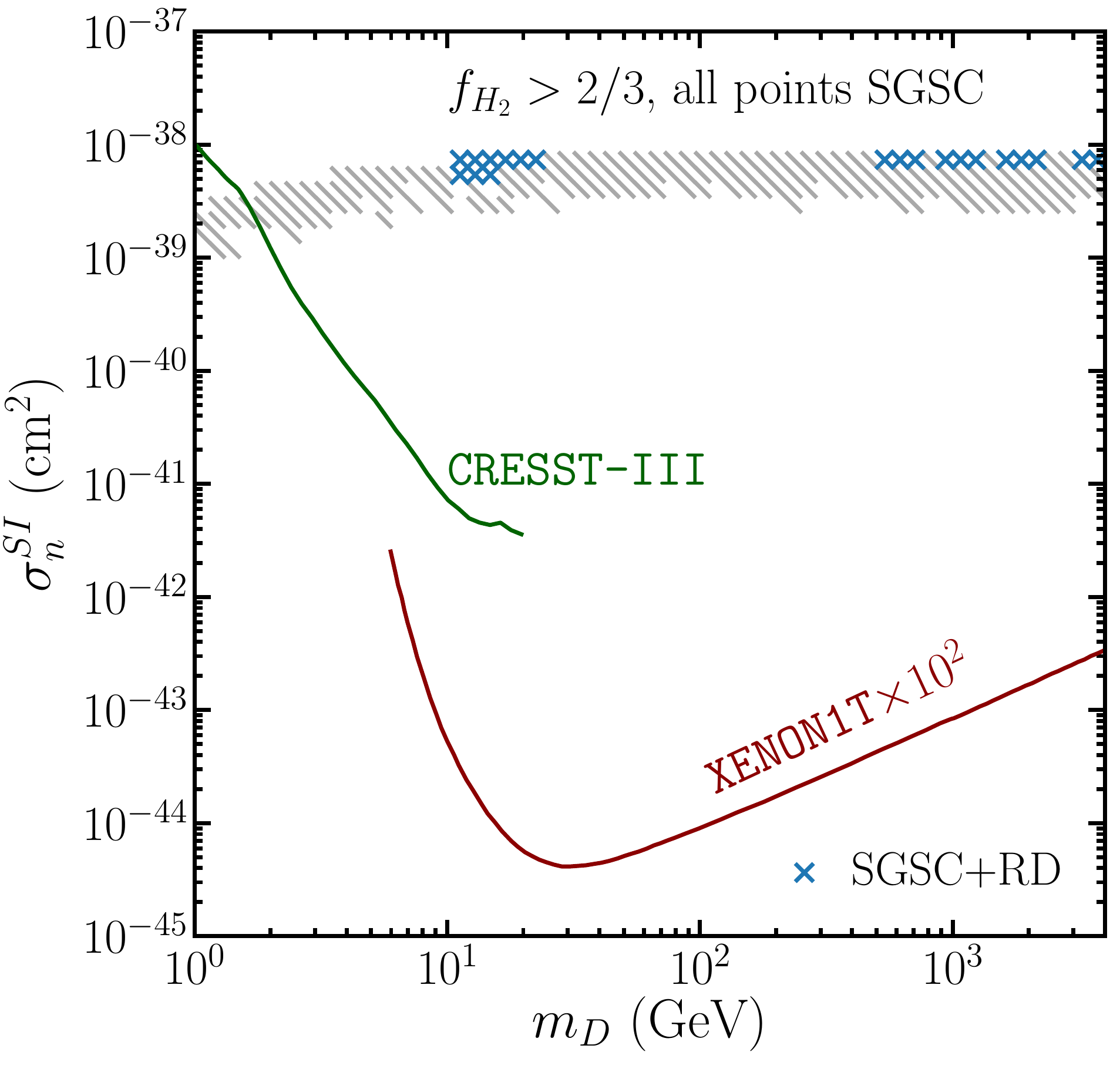}
\caption{
	Doublet-like DM \textbf{SGSC} allowed regions projected on ($m_D$, $\oh$) (left)
	and ($m_D$, $\sigma^{SI}_{n}$) (right) planes.  
	The gray area in the left panel has no coannihilation or resonance. 
         The gray area in the right panel is excluded by PLANCK data at $2\sigma$.
} 
\label{fig:doublet}
\end{figure}

The doublet-like DM in G2HDM is similar to the IHDM case
in the limit where the scalar $(S)$ and pseudo-scalar $(P)$ components 
in $H^0_2$ are mass degenerate.  We show the scatter plot for 
the relic density dependence on the DM mass in
the left panel of Fig.~\ref{fig:doublet}.  
Similar to Refs.~\cite{Goudelis:2013uca,Hambye:2009pw}, 
there are several different annihilation mechanisms governing different DM mass regions. 
However, the observed relic abundance $\oh\approx 0.1$ only occurs 
at around $m_D\sim 10\gev$ and $m_D>500\gev$. 

In the following, we discuss in more detail the DM annihilations for this inert doublet-like DM case 
in the four DM mass regions (i) to (iv) consecutively.
\begin{itemize}
\item[(i)]
First, the DM masses that lies between $1\gev$ to $10\gev$ whose
major contributions of the DM annihilation cross section are given by $DD^*\to\
c \bar{c}$ and $\tau^+ \tau^-$ via $s$-channel SM Higgs exchange.  Despite of the
small  $c$ and $\tau$ Yukawa couplings, the cross section can be slightly
enhanced by the relatively big $DD^*h_i$ coupling, as given in
Eq.~\eqref{eq:g_d_ddh}.  Thanks to large values for $\lambda_{H\Phi}v_\phi$
and $\lambda_{H\Delta}v_\Delta$.  However, the total cross section is too
small to bring the relic density in this mass range closer to
the PLANCK measurement.  Due to the opening of the $DD^*\to b\bar{b}$ channel 
and its larger Yukawa coupling, the correct relic density can
be obtained for DM masses between 11~GeV and 20~GeV.

\item[(ii)]
When DM mass is around half the $Z$ mass, the $Z$ exchange diagram becomes
very efficient and the enhancement in resonant annihilation brings the relic
density well below 0.12 of the PLANCK measurement.  We note that $DD^*Z_i$
couplings are unique in G2HDM due to the nature of complex scalar. In IHDM,
the DM can be either the real or imaginary parts of $H_2^0 = S + i P$, in
which case neither the $SSZ$ nor $PPZ$ coupling is present.  Similarly, it
also happens for DM mass at around half the SM Higgs mass, where again,
enhanced annihilation rate through Higgs exchange brings the relic density
even lower.

\item[(iii)]
If DM mass is increased above half the SM Higgs mass ($m_{h_1}/2<m_D<500\gev$),
the Higgs resonance is no longer efficient.  However, once the gauge boson
final state, especially $W^+W^-$, opens ($m_D>m_W$), the total cross section is
governed by the process $DD^*\to W^+W^-$.  The relevant diagrams for this
process are the 4-point interaction $DD^*W^+W^-$, $s$-channel mediated by each
$h_i$ and each $Z_i$ gauge bosons, and $t$- and $u$-channels with charged
Higgs mediator.  The dominant channel is the $s$-channel through lightest
Higgs $h_{1}$ (not efficient but non-negligible) and second lightest Higgs $h_{2}$ exchange.
The third Higgs $h_3$ is too heavy and not relevant.
Thus the annihilation cross section is determined by the $D D^* h_{1}$ 
and $D D^* h_{2}$ couplings.  
These two couplings have terms proportional to
each of the three VEVs (see Eq.~\eqref{eq:g_d_ddh}), 
which are usually too large in order to have enough relic density at the DM mass region
below 500~GeV.
As a result, the observed relic density (within 2$\sigma$
region) can not be satisfied in this regime as one can see in the left panel
of Fig.~\ref{fig:doublet}. 
The other three final states opening in this intermediate mass range are
$ZZ$, $h_1 h_1$ and $t \bar t$, which are all sub-dominant compared with the $W^+W^-$
final state.

\item[(iv)]
Finally, in the heavy mass region ($m_D>500\gev$), the dominant final states are
from the the longitudinal components of the gauge bosons, namely $W^{+}_{L}W^{-}_{L}$ and 
$Z_{L}Z_{L}$. 
For $Z_{L} Z_{L}$ final state, there is an exact cancellation between the 4-point contact interaction 
diagram and the $t$ and $u$-channels of $D$ exchange diagrams. 
The sum of these 3 diagrams is proportional to 
$(s+t+u - 2m^{2}_{D} - 2m^{2}_{Z})$ and hence vanishes identically due to kinematical constraint. 
Thus the remaining diagrams for $DD^* \to Z_L Z_L$ are given by the $s$-channel $h_{i}$ exchange 
which lead to $S$-wave total cross section in the non-relativistic limit.
There is a similar cancellation between the 4-point contact interaction diagram and the 
$t$-channel charged Higgs exchange diagram for the $W^{+}_{L}W^{-}_{L}$ final state.
The sum of the amplitudes from these two contributions is given by
%
\begin{equation}
\label{eq:wlwl}
\mathcal{A}_{\text{(4-pt + Charged \,Higgs)}} \approx
\frac{e^{2}
(\mathcal{O}^{D}_{22})^{2} }{2 m^{2}_{W} s^{2}_{W}} \left[ \frac{(s-2
m^{2}_{W})}{2} + \frac{(t-m^{2}_{D})^{2}}{(t-m^{2}_{H^{\pm}})} \right] \, ,
\end{equation}
where $t = m^{2}_{D} + m^{2}_{W} - s/2$.
Clearly, when $s$ is sufficiently large such that all masses can be ignored
and $t \sim -s/2$, the above amplitude vanishes.
However one notes that if $D-H^\pm$ coannihilation happens for this heavy DM
mass region, {\it i.e.}, when $m_D \simeq m_{H^{\pm}}$, the above amplitude is
also vanishing.
Thus in the heavy DM mass region where the $D-H^\pm$ coannihilation occurs, 
the dominant diagrams that contribute to $DD^* \to W^{+}_{L}W^{-}_{L}$ 
are the $h_{i}$ and $Z_{i}$ exchanges which give rise to 
$S$-wave and $P$-wave total cross sections respectively in the non-relativistic limit.
We can also conclude that the total cross sections for DM annihilation into 
both $W^+_LW^-_L$ and $Z_LZ_L$ final states in G2HDM
are consistent with unitarity~\cite{Griest:1989wd}.

\end{itemize}

In the right panel of Fig.~\ref{fig:doublet}, we show the scatter plot for the 
spin independent direct detection cross section versus the DM mass.
The interactions between DM and nucleons are mediated by $t$-channel $h_i$ and
$Z_i$ boson exchange, with a small contribution from $u$-channel heavy fermion
exchange. Due to the $SU(2)_L\times U(1)_Y$ charge of the inert doublet $H_2$, the
doublet-like DM-nucleon cross section is dominated by $Z$ exchange.
As one can see in the plot, the doublet-like
DM in G2HDM predicts a typical value of the cross section of order $10^{-38}$~cm$^2$. It can
be excluded by XENON1T~\cite{Aprile:2018dbl} and
CRESST-III~\cite{Petricca:2017zdp} down to DM masses above 2 GeV.
For the points below 2 GeV that survive the CRESST-III constraint, 
the predicted relic abundance is always higher than the measured PLANCK value.  
Regarding the ISV effects, we check that $|f_n/f_p|$ remains typically 3 orders of
magnitude far away from the maximal cancellation value of $f_n/f_p\approx-0.7$. 
Therefore, there is no noticeable reduction in the nucleon-level {\bf DD} cross section.

It is clear from the previous discussion that for the doublet-like DM case 
there is no surviving parameter space 
that can remain after the constraints from both PLANCK and XENON1T are taken into account. 
Therefore, doublet-like DM in G2HDM is completely ruled
out by current experiments, at least under the somewhat generic
conditions set up in this paper. A study of particular mechanisms or very
specific sets of parameters ({\it e.g.} a very light mediator region) 
that may bring down the relic density for light
doublet-like DM ($\sim 1$ GeV) while keeping the prediction of direct detection intact
is out of the scope of the present analysis.


\subsubsection*{$SU(2)_H$ Triplet-like DM}

\begin{figure}[t]
\centering
\includegraphics[width=0.48\textwidth]{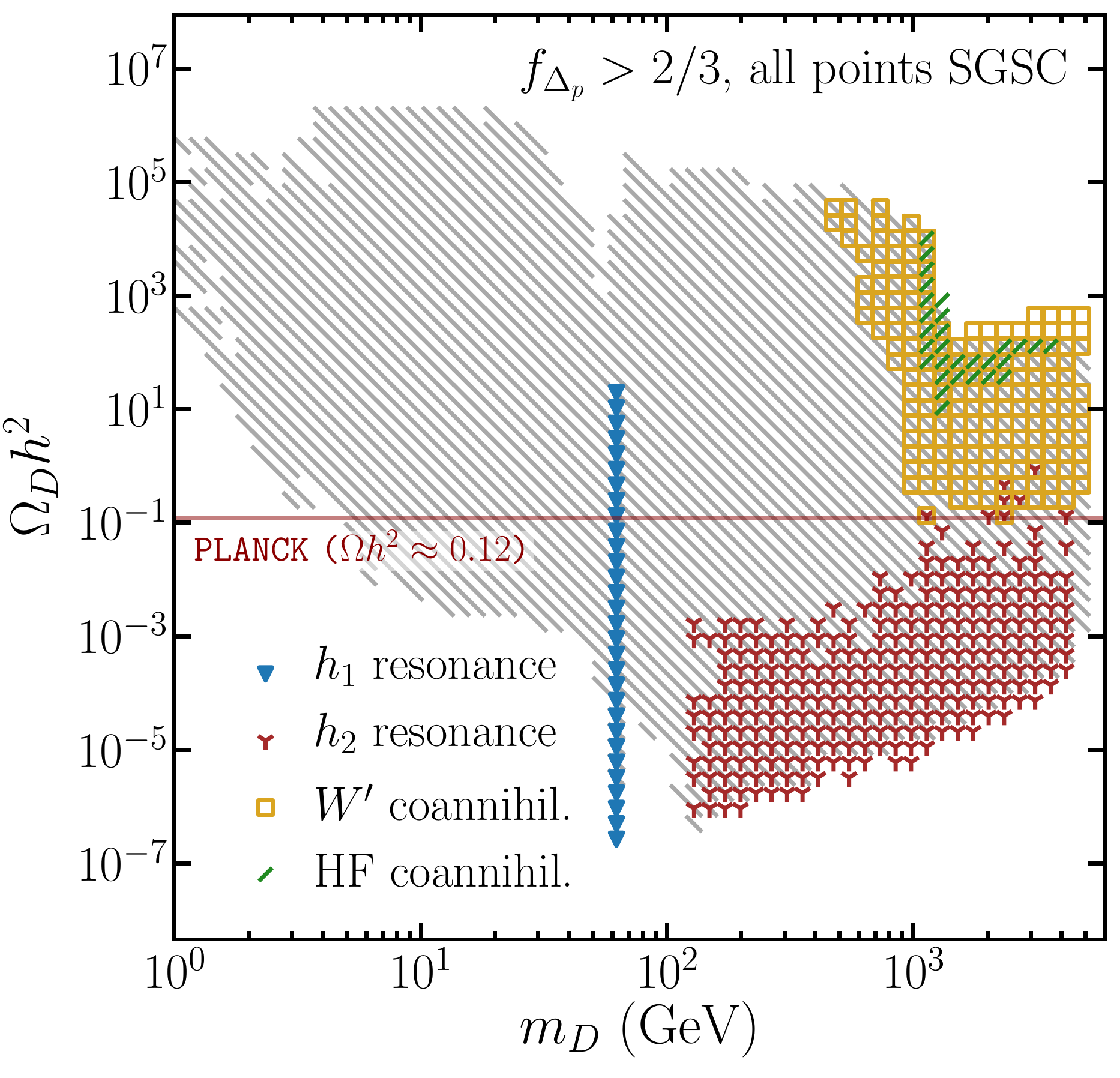}
\includegraphics[width=0.49\textwidth]{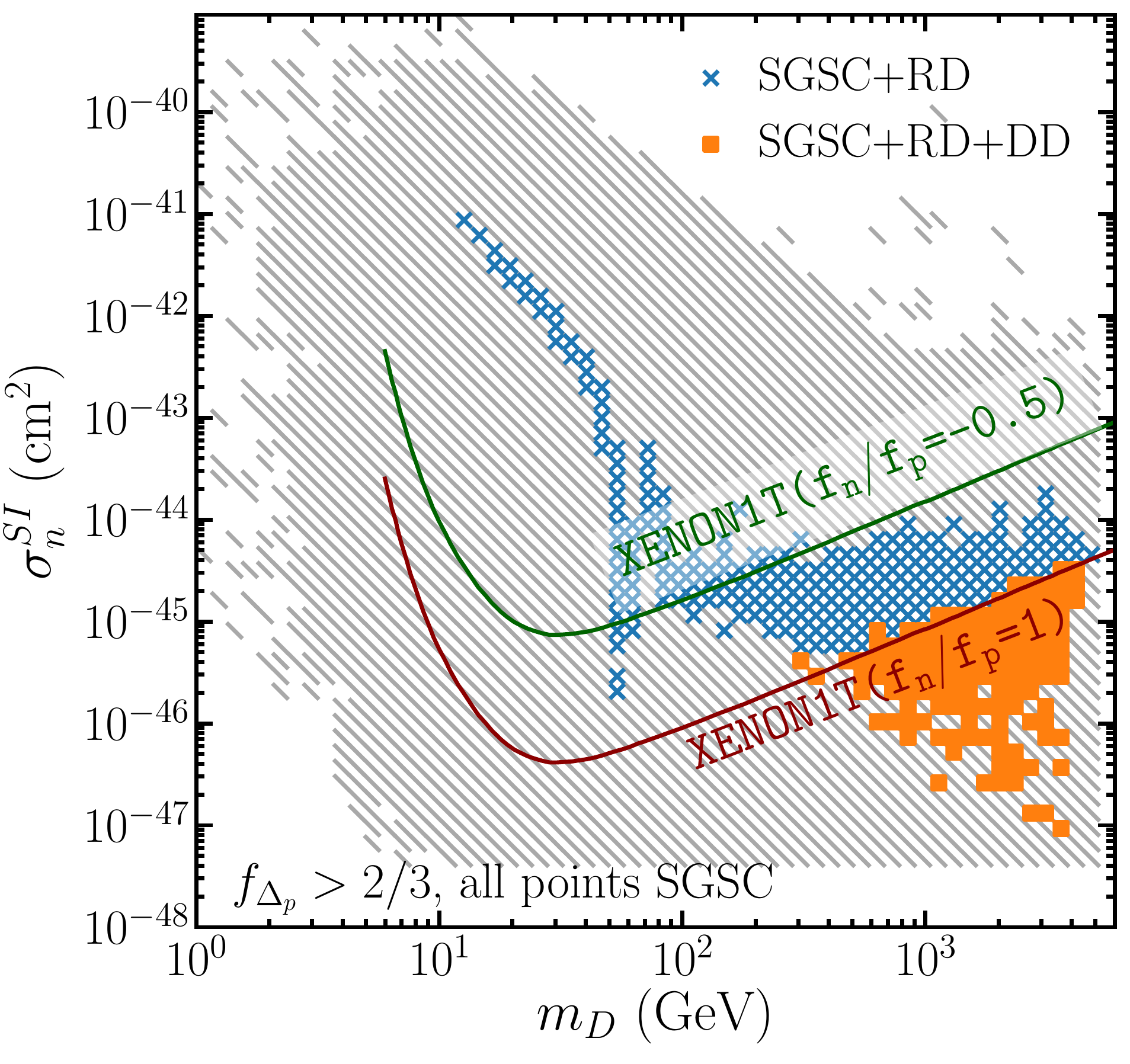}
\caption{
Triplet-like DM {\bf SGSC} allowed regions projected on ($m_D$, $\oh$) (left)
and ($m_D$, $\sigma^{SI}_{n}$) (right) planes.  The gray area in the left panel
has no coannihilation or resonance.  The gray area in the right panel is
excluded by PLANCK data at $2\sigma$. 
In the right panel, the lower red solid line is the published XENON1T limit with isospin
conservation, while the upper green solid line is the same limit but for
ISV with $f_n/f_p=-0.5$.
Some orange filled squares are above the published XENON1T
limit due to ISV cancellation at nucleus level.
} 
\label{fig:triplet}
\end{figure}

One fundamental difference between triplet-like and doublet-like
DM is that now $D$ is dominated by the term $\mathcal{O}^D_{32}\Delta_p$ in
Eq.~\eqref{eq:Dcomposition}.
Therefore, one should expect all the couplings to behave differently from 
the previous doublet-like case. In particular, the coupling terms that were
relevant for doublet-like DM will now be suppressed by a smaller $\mathcal{O}^D_{22}$.
In Fig.~\ref{fig:triplet}, we show the scatter plots for the relic density and spin independent 
direct detection cross section versus the DM mass $m_D$ 
at the left and right panels respectively for the triplet-like case. 
Similar to the doublet-like case, one can divide the DM mass in different regions for discussions.
The opening channels are the same in each region and hence no need to repeat here.
However the dominant channels in each region
may be changed due to the differences of the couplings in both DM compositions.

For the relic density, the resulting resonances and coannihilation regions are presented 
in the left panel of Fig.~\ref{fig:triplet}.
We found that in the DM mass range below $m_{h_1}/2$ (region (i)), the dominant DM
annihilation contribution to the relic abundance comes from $s$-channel Higgses exchange
with final states of $\tau^+\tau^-$ and $b\bar{b}$. 
By looking at the $DD^*h_i$ coupling in 
Eq.~\eqref{eq:g_T_ddh}, it is easy to see that the large value
of $v_\Phi$ makes annihilation through $h_2$ ($\delta_3$-like) comparable with
annihilation through $h_1$ (SM-like) while the heavier $h_{3}$ ($\phi_2$-like)
contribution remains subleading. As expected, the lowest
relic density happens at the Higgs resonance region $m_D\approx
m_{h_1}/2$, combining with the large $DD^*h_1$ coupling from the first term of
Eq.~\eqref{eq:g_T_ddh}.  Of course, one can always decrease the values of
$v_\Phi$ or $v_\Delta$ to reduce the coupling size for larger relic abundance, 
but this is not particularly interesting for a thermal DM scenario.
As DM becomes heavier, other resonance turns on. For $m_D > 100$ GeV, $D$ is
massive enough to have points where $2m_D \approx m_{h_2}$ and some points
resulting in resonant annihilation through $s$-channel $h_2$ exchange. 
In contrast to the doublet-like scenario, it is possible for the triplet-like DM to have a very
wide range of relic density values, given the several different possible combinations 
for the $DD^*h_i$ coupling in Eq.~\eqref{eq:g_T_ddh}.
Unlike the doublet-like case, for the triplet-like DM case in region (ii), the reduction 
of the relic density due to 
the $Z$ resonance enhancement in the annihilation cross section 
is absent because the triplet $\Delta_H$ is a SM singlet, 
meaning that the interaction between DM
and the SM $Z$ is suppressed by product of small mixing elements like
$(\mathcal{O}^D_{32})^2 \mathcal O^G_{21}$ according to Eq.~\eqref{eq:g_T_ddz}.

In region (iii) where $m_{h_1}/2 < m_D < 500$ GeV, the relic density reduction mechanism is
similar to the doublet-like DM case discussed above. The annihilation cross
section is highly dominated by $W^{+}W^{-}$ (more than $\sim 50 \%$),
$h_{1}h_{1}$ ($\sim 25\%$), and $ZZ$ ($\sim 20\%$) final states.  
The main contribution to the $W^{+}W^{-}$ final state
comes from $S$-wave given by $h_i$ exchange, while the
$P$-wave contribution is suppressed and originated from neutral gauge bosons
mediator exchange.
The $S$-wave annihilation cross section is
controlled by the $D D^* h_{1}$ and $D D^* h_{2}$ couplings, as can be
seen in Eq.~\eqref{eq:g_T_ddh}.  The contribution from $h_{3}$ exchange is
negligible because of its heavy mass.  

In region (iv) where $m_D > 500$ GeV, DM annihilates into $W^+_LW^-_L$ predominantly
while other channels are subdominant, similar to the doublet-like DM case. There is no need to elaborate further 
here.

Generally speaking, the charged Higgs $H^\pm$ contribution here can be omitted since
it is more than twice heavier than the DM $D$. Differently from the doublet-like
case, there is no coannihilation between $H^{\pm}$ and $D$ in the triplet-like
DM case.  Next, the coannihilation between DM and $\widetilde{\Delta}$ is
absent as well because the $\widetilde{\Delta}$ is also much heavier than $D$
due to the choice of larger $v_\Delta$ to make the (3,3) entry of
Eq.~\eqref{eq:Z2oddmassmatrix} smaller.  Therefore, the only possible
efficient coannihilation is between DM and $W^\prime$ for DM mass above 400~GeV
(orange boxes at the left panel of Fig.~\ref{fig:triplet}).
This coannihilation is only important  for relic density above $0.12$, where
some $DD^*$ annihilation channels may be insufficient because their couplings
to $h_i$ and $Z_j$ may be suppressed.  A small region with heavy fermion
coannihilation happens for $m_D > 1$ TeV with relic density above 10 
(green shaded points in the left panel of Fig.~\ref{fig:triplet}).  This is
close to the maximal relic density in our scan for that mass range. This
indicates
that heavy fermion coannihilation is important only when the other
annihilation channels are strongly suppressed.

Regarding direct detection, due to the $DD^*Z$ coupling suppression by mixings 
in this triplet-like case,
the elastic DM-nucleon scattering spin independent cross section mediated by the $h_i$ 
and the extra neutral gauge bosons $Z^\prime$ and $Z^{\prime\prime}$
bosons may be relevant. 
We confirm that the dominant contribution to the spin independent cross section is
given by $h_{1}$ and the next dominant contributions are $Z$ and $Z^\prime$, while $h_2$, $h_3$ and
$Z^{\prime\prime}$ are always subdominant.
The contributions mediated by heavy
fermions are negligible due to suppression by their masses in the propagators.

In the right panel of Fig.~\ref{fig:triplet}, we can see that for $m_D\gtrsim 300$ GeV it is possible
to find a region that agrees with relic density constraint from PLANCK at
$2\sigma$ and remains below the published XENON1T limit 
at the neutron with $f_n/f_p = 1$. Note that some
of the allowed points (orange squares) are above this XENON1T limit.
This is due to mild ISV cancellation that brings such points below
the XENON1T limit at nucleus level, as given by Eq.~\eqref{eq:sigmaEXP}.
For comparison, the XENON1T limit at the neutron level 
with ISV of $f_n/f_p = -0.5$ is also shown.

\begin{figure}[!htb]
\includegraphics[width=0.48\textwidth]{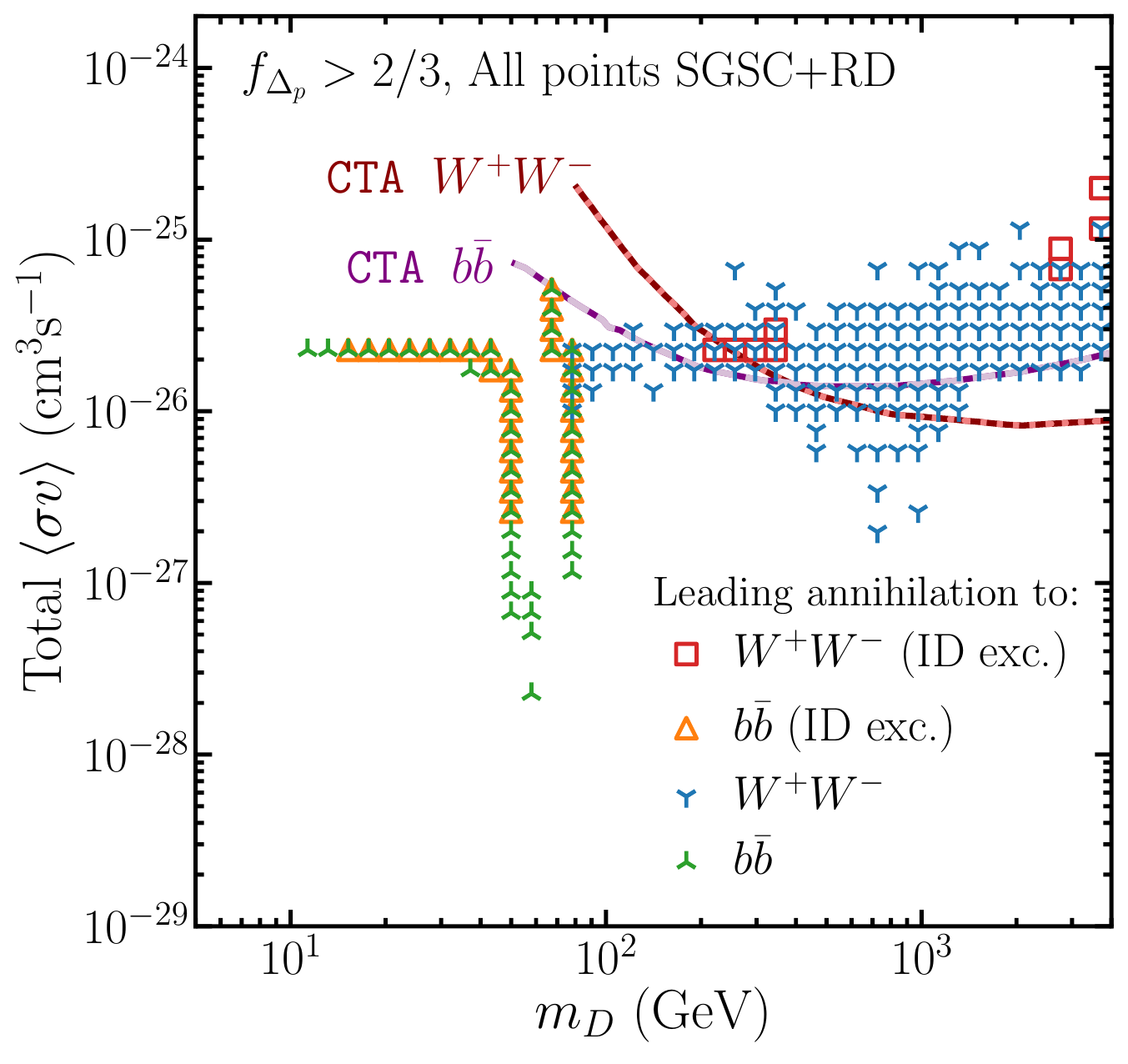}
\includegraphics[width=0.48\textwidth]{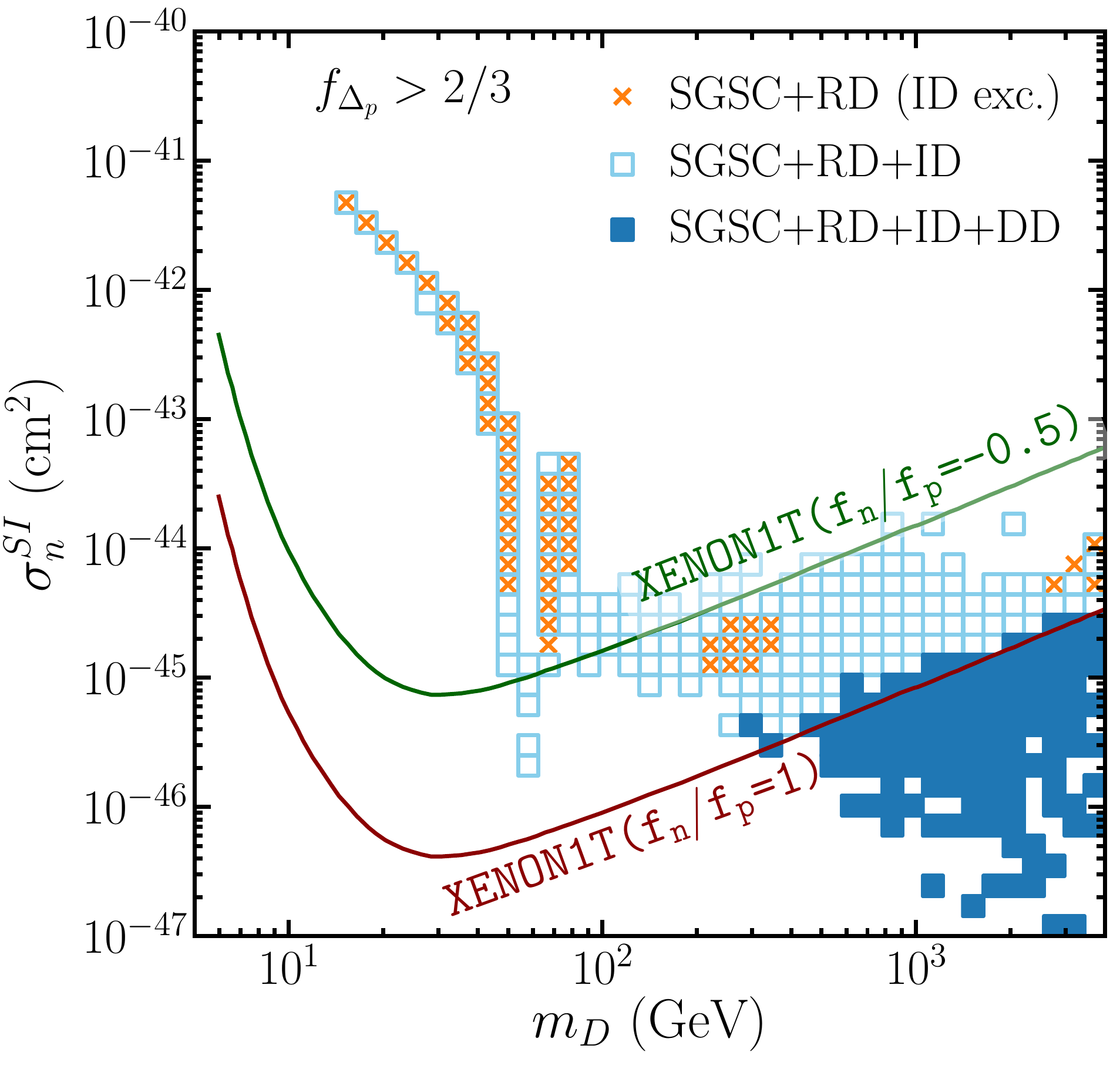}
\caption{
The present time total annihilation cross section according to dominant
annihilation channel (left) and DM-neutron elastic scattering cross-section
(right) for $f_{\Delta_p}>2/3$ in the triplet-like DM case versus the DM mass $m_D$.  
Two-dimensional $2\sigma$ criteria of the ID constraints is $\Delta\chi^2=5.99$ 
based on Fermi dSphs gamma-ray flux data.
Future CTA measurements may
help constrain regions with DM masses above $\mathcal{O}(10^2)$~GeV as shown
in the left panel.
In the right panel, the lower red solid line is the published XENON1T limit with isospin
conservation, while the upper green solid line is the same limit but for
ISV with $f_n/f_p=-0.5$.
Some blue filled squares are above the published XENON1T
limit due to ISV cancellation at nucleus level.
\label{fig:TLID}
}
\end{figure}

The constraint of indirect detection from Fermi-LAT's gamma-ray observation
imposed on the triplet-like DM is shown in Fig.~\ref{fig:TLID}.  The left
panel presents the DM annihilation cross section dependence
\footnote{
Note that to apply Fermi-LAT constraints we use photon flux as
calculated with Eq.~\eqref{eq:idflux}.
The annihilation channels displayed in Fig.~\ref{fig:TLID} are
only leading channels that may not be significantly above other channels.
}
on the DM mass at the present universe with \textbf{SGSC+RD}.
Results are only presented for the dominant annihilation channels, 
$b \bar b$ and $W^+W^-$. 
One can see that DM with $m_D\lesssim 90\gev$ mainly annihilates to
$b\bar{b}$. 
At the region near the $Z$ or $h_{1}$ resonance, 
the corresponding cross section at the 
present universe drops while satisfying the relic density. 
This is a typical feature of the resonance region because 
the DM relative velocity at the early universe is much larger than the value at the present 
one.
In order to cancel a large cross section caused
by the resonance at the early universe,
a small coupling of $DD^*Z$ or $DD^* h_1$ is required to
make $\sigmav$ at the early universe comes close to the canonical value of 
$10^{-26}$~cm$^3 \cdot$s$^{-1}$.
However, when the universe temperature drops, the resonance cannot be
maintained by the kinetic energy of DM at the present day. At this time the
cross section becomes smaller and is hard to be observed by 
Fermi-LAT.

Once DM mass is heavier than $W^\pm$ boson mass, the final state
$W^+ W^-$ starts dominating the annihilation cross section rapidly.
Note that the current  \textbf{ID} sensitivity can only apply strongly for the DM mass
located between 10~GeV and few hundred GeV.  However, the future CTA
sensitivity~\cite{Morselli:2017ree} might reach the TeV region of $m_D$ and
further constrain our parameter space, as show in the left panel of
Fig.~\ref{fig:TLID}.

In the right panel of Fig.~\ref{fig:TLID}, we
display the exclusion from \textbf{ID} projected on the plane of DM-neutron spin
independent cross section $\sigma_n^{SI}$ 
versus $m_D$. We can see that all the \textbf{ID} excluded points sit
above the limit set by XENON1T. The exclusion limits are 
given by recent XENON1T data (blue unfilled squares) and Fermi gamma-ray constraints
(orange crosses).  One can see the XENON1T exclusion power
is much stronger than Fermi gamma-ray exclusion. 


\subsubsection*{$SU(2)_H$ Goldstone boson-like DM}
\label{sec:Goldstone}

\begin{figure}[htb]
\centering
\includegraphics[width=0.48\textwidth]{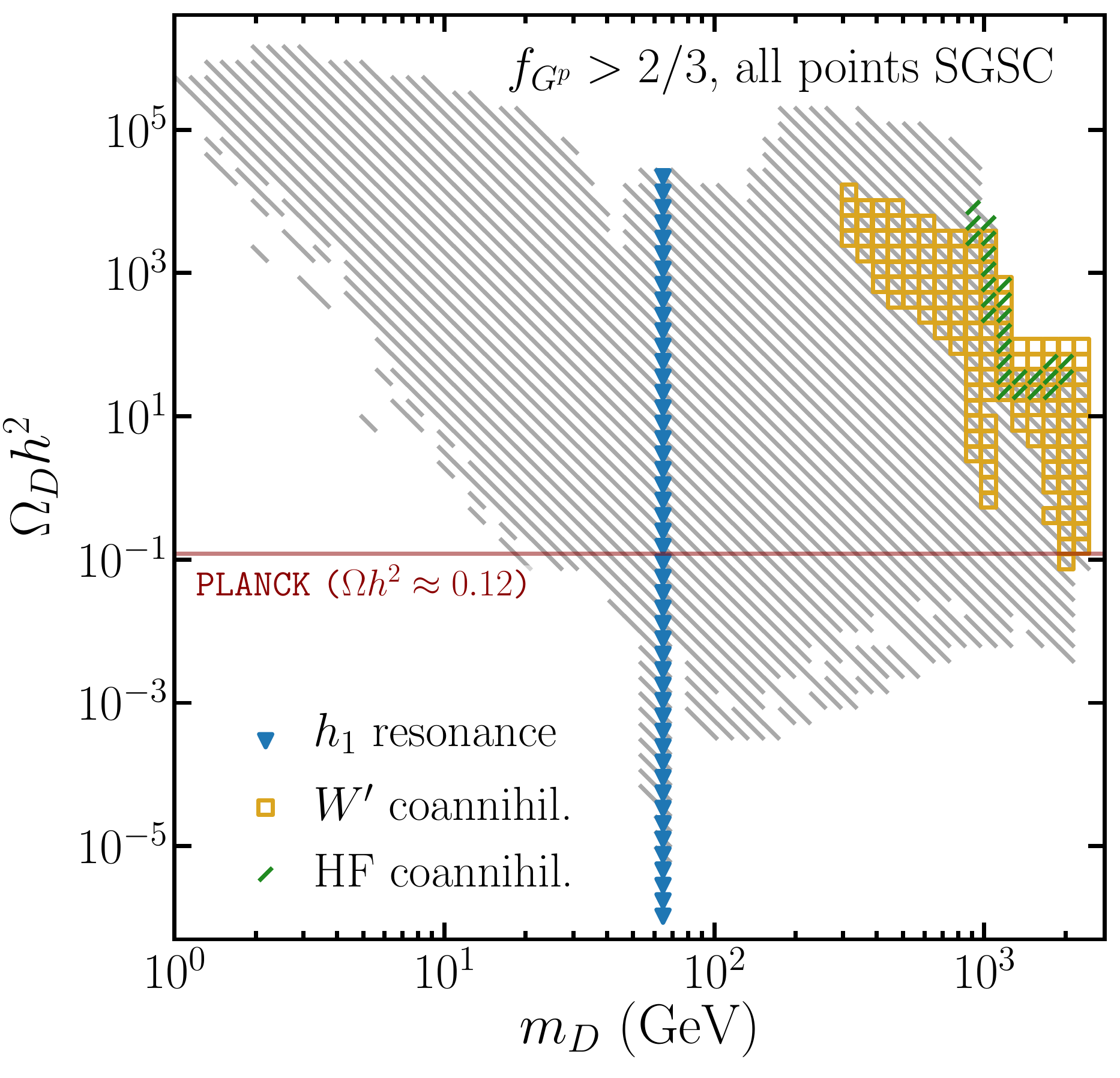}
\includegraphics[width=0.49\textwidth]{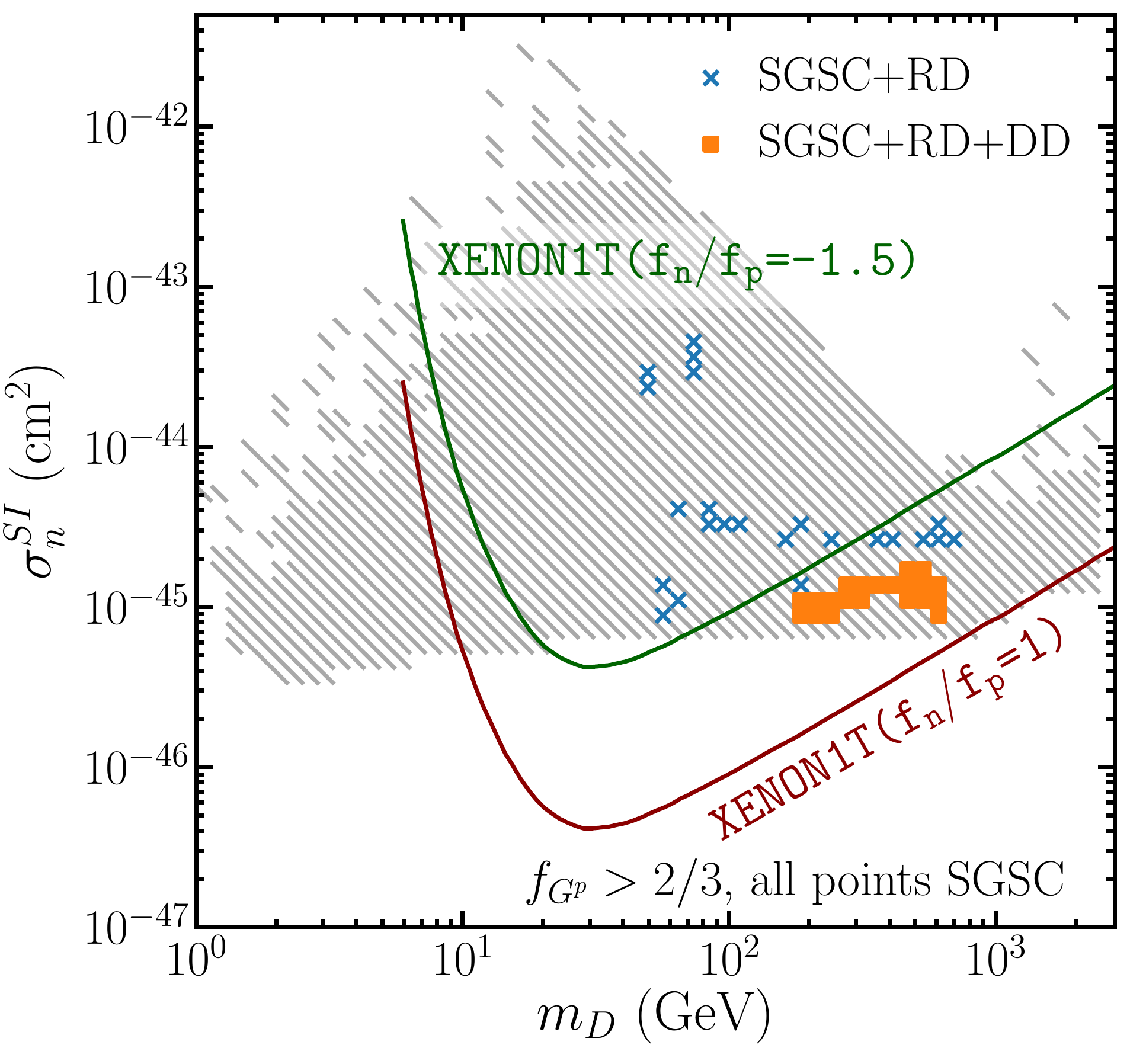}
\caption{
	Goldstone-like DM {\bf SGSC} allowed regions projected on ($m_D$, $\oh$) (left)
	and ($m_D$, $\sigma^{SI}_{n}$) (right) planes.  The gray area in the left panel
	has no coannihilation or resonance.  The gray area on the right is
	excluded by PLANCK data at $2\sigma$. 	
	In the right panel, the lower red solid line is the published XENON1T limit with isospin
	conservation, while the upper green solid line is the same limit but for
	ISV with $f_n/f_p=-1.5$. 
	The small region of orange filled squares above the published
	XENON1T limit present ISV cancellation at nucleus level. 
 }
\label{fig:goldstone}
\end{figure}

In this case, as shown at the end of Sec.~\ref{sec:HiddenP} (see Fig.~\ref{fig:vD_vP_fgp}), 
the Goldstone-like DM $D$ will be a mixture dominated by $G^p_H$
with an important component coming from $\Delta_p$, while the $H^{0*}_2$ 
component remains suppressed. 
In the left panel of Fig.~\ref{fig:goldstone}, 
for the DM mass regions (i)-(iii), the dominant channels for DM annihilation 
in the relic abundance calculation are similar to the triplet-like DM case.
In heavy mass region (iv), the cross section 
is again dominated by the $W^{+}_{L}W^{-}_{L}$ final state
which contributes $\sim$50\%, while the transverse component is negligible.
The main difference between Goldstone-like and triplet-like DM 
can be understood by their corresponding dominant couplings. 
For triplet-like DM, the dominant couplings are given by Eqs.~\eqref{eq:g_T_ddz} and \eqref{eq:g_T_ddh} which are proportional to the $(\mathcal{O}^{D}_{32})^{2}$ characterizing the corresponding $\Delta_p$ component. Similarly, one expects that the Goldstone-like DM receives its dominant couplings purely via Eqs.~\eqref{eq:g_G_ddz} and \eqref{eq:g_G_ddh}. However, this is not the case for Goldstone-like DM. 
There is also an important contribution coming from the $\Delta_p$ part in the relevant couplings. Thus, one needs to include not only  the couplings proportional to $(\mathcal{O}^{D}_{12})^{2}$ but also the ones proportional to $(\mathcal{O}^{D}_{32})^{2}$.
The effect of $\Delta_p$ component in this case is reducing the $DD^{*}h_{1}$ and $DD^{*}h_{2}$ couplings while enhancing the $DD^{*}Z$ and $DD^{*}Z'$ couplings. As a consequence, the dominant DM annihilation channel $W^{+}_{L}W^{-}_{L}$ will be dominated by $P$-wave component originated from the $Z^\prime$ exchange, while the $S$-wave part coming from the $h_1$ and $h_2$ mediators is subdominant. 
The next important contribution is given by the $Z' Z'$ final state.
The $Z' Z'$ final state occurs via four point contact interaction, $t$ and $u$-channels of $D$ exchange and $s$-channel of neutral Higgses exchange. The presence of  the $\Delta_p$ component in the Goldstone-like DM further enhance the $DD^{*}Z'Z'$ coupling resulting the appearance of the new important final state $Z' Z'$ 
in the heavy mass region (iv).

%

%



Coannihilation in this case is very similar to the triplet-like DM case. The
most relevant coannihilations happen with $W^\prime$ and heavy fermions for large
masses and large relic density. Coannihilation with $W^\prime$ only presents when 
the DM mass gets close to 300~GeV and its relic density mostly above the PLANCK measurement. 
As the triplet-like DM case, 
the usual $DD^*$ annihilation channels become smaller leaving more
way for coannihilations that, otherwise, would be negligible. For the case of
heavy fermions, coannihilation happens for DM masses above 1~TeV and mostly
for the upper bound of relic density, where $DD^*$ coannihilation is even
more suppressed than for the $W^\prime$ case.


In the right panel of Fig.~\ref{fig:goldstone} we show the scatter plot for the DM-neutron cross
section dependence on the DM mass.
The dominant contribution comes from  $h_{1}$ exchange 
with the next dominant ones given by the exchange of $Z$ and $Z'$ bosons. 
The bottom part of the gray region in Fig.~\ref{fig:goldstone} comes mostly from interactions mediated by
the $Z$ and $Z'$ gauge bosons and is limited from below by our lower limit for
the scan range of $g_H$ determined by Eq.~(\ref{eq:ghmincondition}).
The interference between $h_1$, $Z$, and $Z^{\prime}$ exchange makes the spin
independent cross section varies in a wide range.
The orange points located between $200 \text{~GeV} \leq m_{D} \leq 600 \text{~GeV}$ satisfy the
observed relic density while escaping the current bound on direct detection
given by XENON1T experiment. The dominant contribution for these points is
given by the gauge bosons exchange $Z$ and $Z'$. 
Due to the fine-tuning parameter space for the Goldstone-like DM mentioned earlier,
only the ratio of $f_n/f_p = -1.86$ has enough ISV cancellation to satisfy the published 
XENON1T limit assuming isospin conservation.
For comparison, the XENON1T limit with ISV of $f_n/f_p = -1.5$ is also shown.

\begin{figure}[tb]
\includegraphics[width=0.48\textwidth]{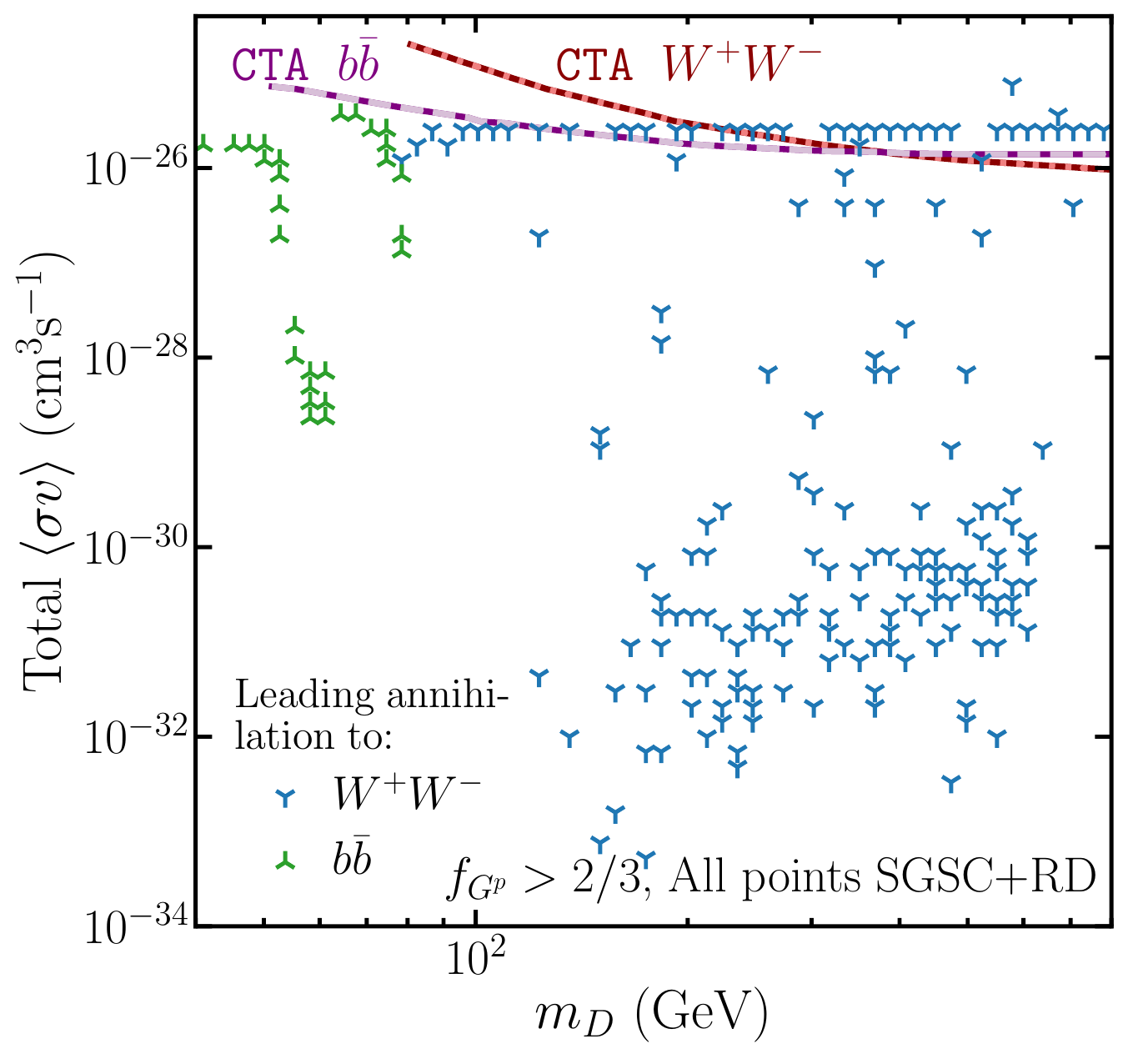}
\includegraphics[width=0.48\textwidth]{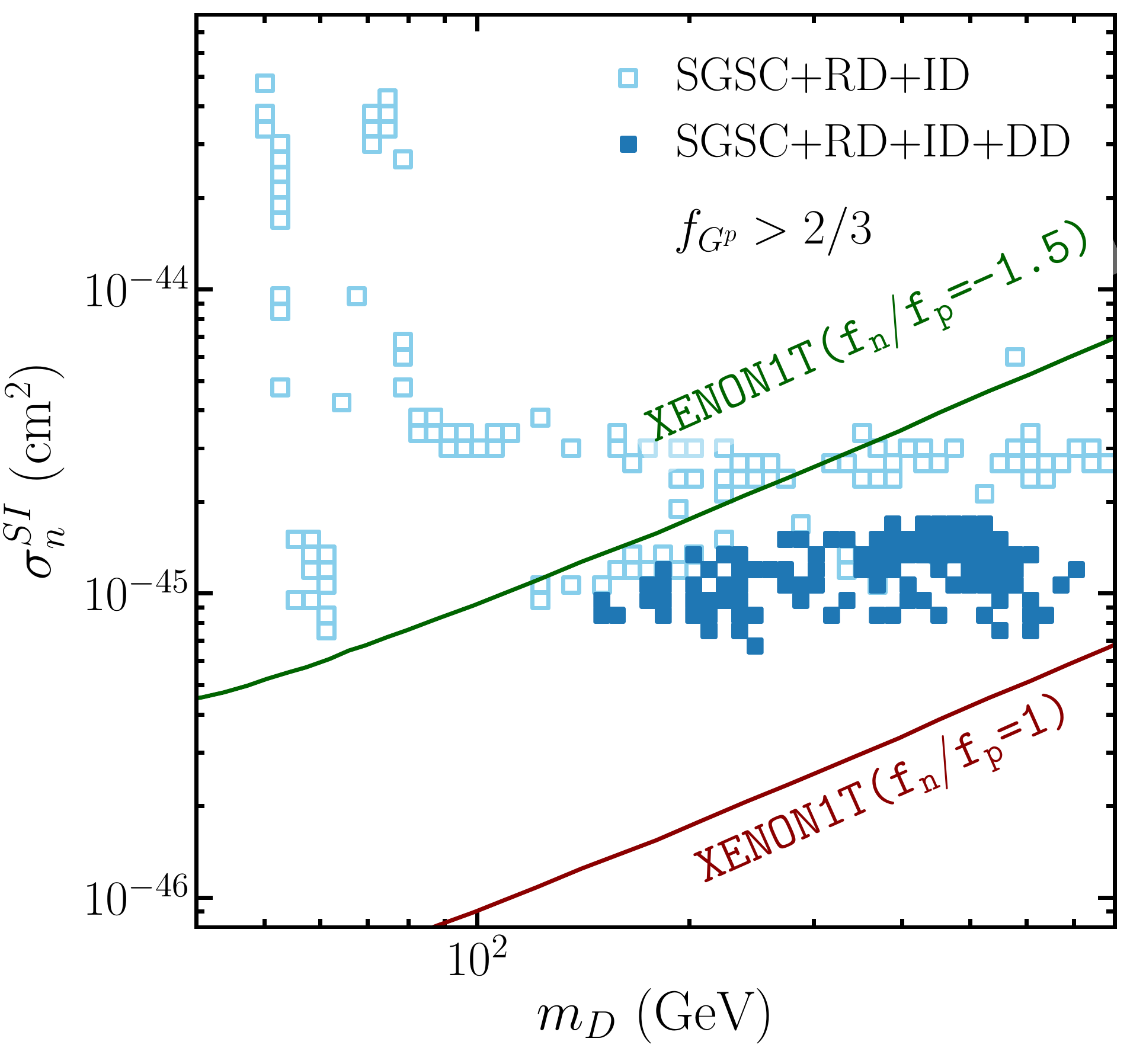}
\caption{
The present time total annihilation cross section by dominant
annihilation channels (left) and DM-neutron elastic scattering cross-section
(right) for $f_{G^p}>2/3$ in the Goldstone-like DM case versus the DM mass $m_D$.  
Two-dimensional $2\sigma$ criteria of the ID constraints is $\Delta\chi^2=5.99$ 
based on Fermi dSphs gamma-ray flux data.
Future CTA measurements may help constrain regions with DM masses 
above $\mathcal{O}(10^2)$~GeV as shown in the left panel.
In the right panel, the lower red solid line is the published XENON1T limit with isospin
conservation, while the upper green solid line is the same limit but for
ISV with $f_n/f_p=-1.5$.
Some blue filled squares are above the published XENON1T
limit due to ISV cancellation at nucleus level.
\label{fig:GLID}
}
\end{figure}

In the \textbf{ID} side, there is no relevant constraining for this Goldstone
boson-like case. Because of $P$-wave suppression of the $Z$ and $Z^\prime$ exchange
in the dominant channels of $b \bar b$ and $W^+W^-$,
most of the points in agreement with the relic density measurement from 
PLANCK have a very low annihilation cross section at the present time and are far
beyond the reach of current experiments of indirect detection,
as can be seen clearly in the zoomed in region on the  $(\langle \sigma v \rangle, m_D)$ plane 
at the left panel in Fig.~\ref{fig:GLID} allowed by the \textbf{SGSC+RD}.
For DM masses below 100~GeV, the
annihilation is dominated by $b\bar{b}$ final state with 90\% of the total
cross section in average. For DM mass above the mass of the $W^\pm$, the
$W^+W^-$ final state dominates completely with an average of 50\% of the total
cross section. 
Unlike triplet-like DM, \textbf{ID} alone does not further constrain the
points allowed by PLANCK. 
The right panel of Fig.~\ref{fig:GLID} shows the zoomed in region of points
on the $(m_D, \sigma_n^{SI})$ plane allowed by the \textbf{SGSC+RD+ID} 
and \textbf{SGSC+RD+ID+DD}.
As mentioned before, ISV effect
($f_n/f_p\approx -1.86$) reduces the sensitivity of the XENON1T result and some
points pass all the constraints ({\bf SGSC+RD+ID+DD}) even though they are
above the direct detection limit at nucleon level. 
Note that there are no points satisfying \textbf{SGSC+RD+DD+ID} beyond 
$m_D \sim 1$ TeV in this Goldstone-like case.

\subsection{Constraining Parameter Space in G2HDM}
\label{sec:ps}

\begin{figure}[!htb]
\includegraphics[width=1.0\textwidth]{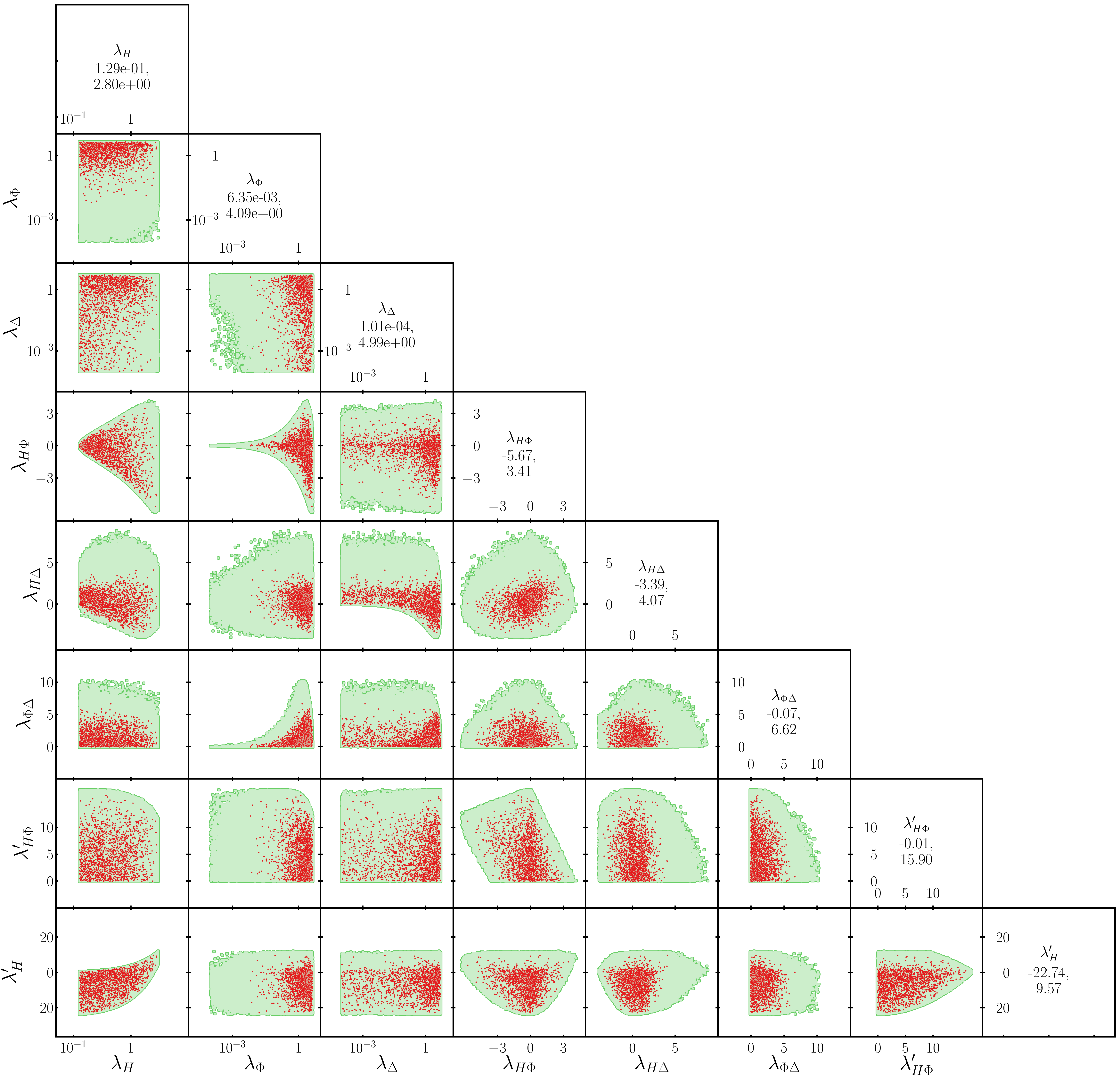}
\caption{A summary plot for the scalar potential parameter space allowed by the \textbf{SGSC} constraints 
(green region) and \textbf{SGSC+RD+DD} constraints (red scatter points) for the triplet-like DM.
The numbers written in the first block of each column are the 1D allowed range of 
the parameter denoted in horizontal axis after the \textbf{SGSC+RD+DD} cut.}
\label{fig:para_lam}
\end{figure}

From previous sections, we have learned that the doublet-like DM scenario
cannot fulfill the DM constraints and that the Goldstone-like DM
requires some fine-tuning in the parameter space and
to escape the XENON1T limit a particular value of $f_n/f_p\approx -1.86$ is required.
Therefore, we will be focusing on discussing the allowed G2HDM parameter
space based on the triplet-like DM. 

In Fig.~\ref{fig:para_lam}, we present the allowed regions of the quartic couplings from the
\textbf{SGSC} constraints (green region) and \textbf{SGSC+RD+DD} constraints
(red scatter points).  Comparing the green regions with the red
scatter points in Fig.~\ref{fig:para_lam}, one can easily obtain the following results:
\begin{itemize}

\item
The allowed ranges on $\lambda_H$ and $\lambda_H^\prime$ remain more or less 
the same before and after imposing \textbf{RD+DD} constraints.

\item
$\lambda_\Phi$, $\lambda_{H\Delta}$, and $\lambda_{\Phi\Delta}$ are mostly constrained by
\textbf{RD+DD} constraints.  To understand this effect, one can see from
Eq.~\eqref{eq:g_T_ddh} that there are three dominant terms that
contribute to the $DD^*h_j$ couplings, $\lambda_{H \Delta} v \mathcal{O}_{11}$,
$\lambda _{\Phi \Delta } v_{\Phi }  \mathcal{O}_{22}$, and $\lambda _{\Delta } v_{\Delta
} \mathcal{O}_{33}$ for $j=1,2,3$ respectively. 
Clearly, $\lambda_{H\Delta}$, and $\lambda_{\Phi\Delta}$ are
restricted by the allowed Higgs coupling sizes. 

\item
Regarding to lighter mediator, in particular for $h_2$, the mixing $\mathcal{O}_{22}$
is strongly related to $\lambda_\Phi$ so that $\lambda_\Phi$ and
$\lambda_{\Phi\Delta}$ are correlated as shown in the third row from
bottom to top and second column of Fig.~\ref{fig:para_lam}. 
These two parameters are related to $h_1$ decay to $f\bar{f}$ and are
constrained by Higgs physics and further by {\bf DD} constraints.

\item
However, $\lambda _{\Delta }$
is not constrained because 
either the DM annihilation or DM-nucleon elastic scattering cross section via the exchange of $h_3$ 
is suppressed by its heavy mass $m_{h_3}$.

\item 
On the other hand, the two off-diagonal terms $\lambda_{H\Phi}$ and
$\lambda'_{H\Phi}$ are constrained mildly.
This is due to the loose requirement that we set for the triplet-like DM
$f_{\Delta_{p}} > 2/3$. In fact, we checked that there can be an important
contribution from the $G^{p}_{H}$ component with $f_{G^{p}_{H}}$ up
to $1/3$.
As a consequence, even though $\lambda_{H\Phi}$ and $\lambda'_{H\Phi}$ do not
appear explicitly in the coupling of $DD^*h_i$ given in
Eq.~\eqref{eq:g_T_ddh}, they appear via subdominant component $G^{p}_{H}$ as
seen in Eq.~\eqref{eq:g_G_ddh}.

\end{itemize}

\begin{figure}[!htb]
\includegraphics[width=1.0\textwidth]{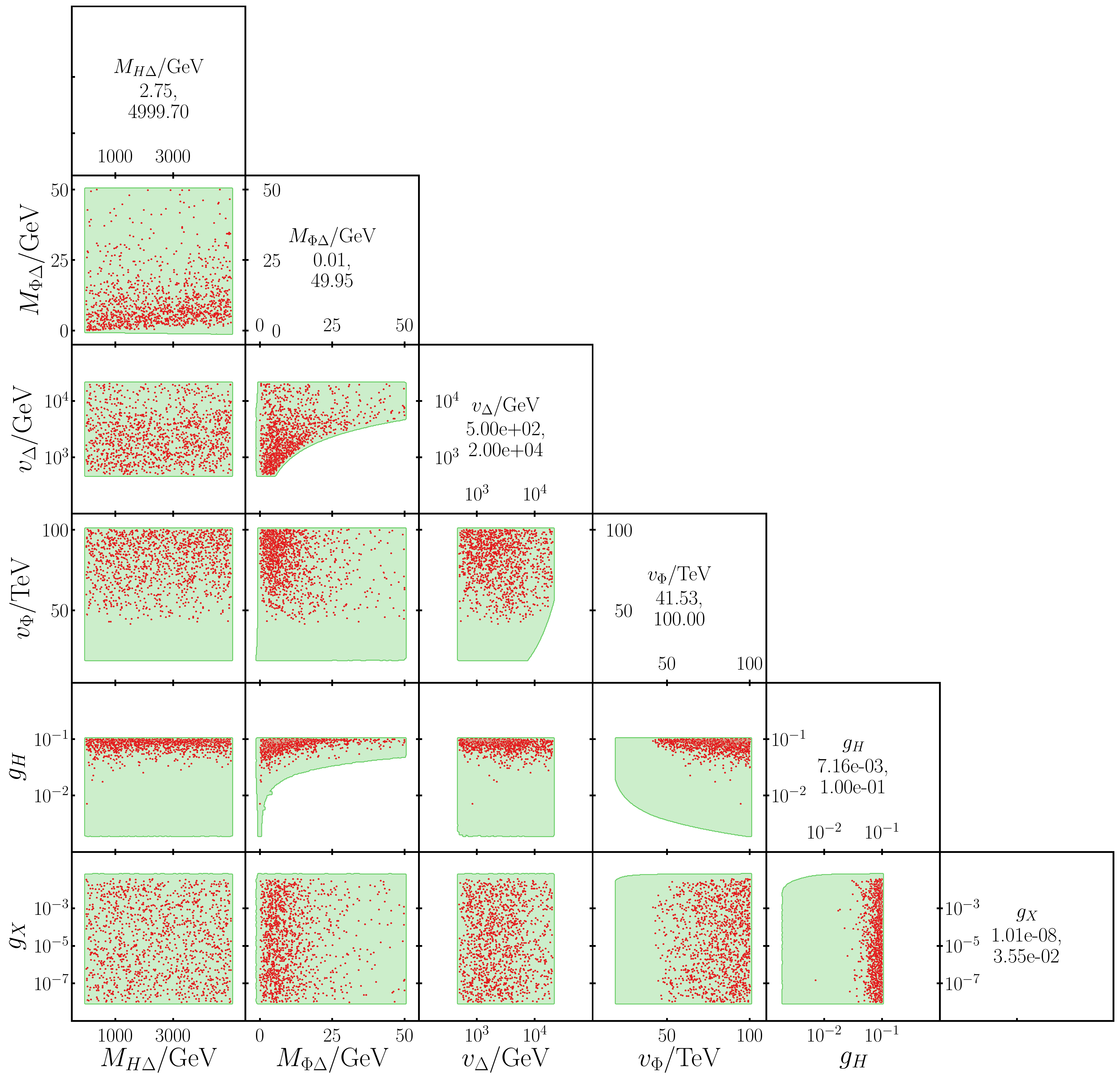}
\caption{
A summary plot for the VEVs, $M_{\Phi\Delta}$, $g_X$ and $g_H$ parameter
space allowed by the \textbf{SGSC} constraints (green region) and
\textbf{SGSC+RD+DD} constraints (red points) for the triplet-like DM.
}
\label{fig:para_gauge}
\end{figure}

Next, we project the allowed G2HDM parameter space to the two VEVs $v_\Phi$ and $v_\De$, 
the two cubic couplings $M_{H\De}$ and $M_{\Phi\De}$, and the two 
new gauge couplings $g_H$ and $g_X$ in Fig.~\ref{fig:para_gauge}. 
Again, by comparing the green regions and the red scatter points in  Fig.~\ref{fig:para_gauge}, 
we can arrive at the following results:
\begin{itemize}

\item
Strikingly, only $g_H$ and $v_\Phi$ can be further constrained by \textbf{RD+DD}. 
Interestingly, we found such an exclusion comes from the lower allowed DM mass. The allowed DM mass
values range from hundreds of GeV to a few TeV.  This range is reflected in $g_H$
since the minimal value we choose for $g_H$ is given by
Eq.~\eqref{eq:ghmincondition} and depends directly on the DM mass.

\item
The other 4 parameters $g_X$, $v_\De$, $M_{H\De}$ and $M_{\Phi\De}$ are not sensitive to the 
dark matter physics constraints from \textbf{RD+DD}.
\end{itemize}

In summary, given the setup of the parameter space in our numerical scanning,
a good WIMP candidate in G2HDM is the triplet-like complex scalar with a mass $m_D$ in the 
electroweak scale, and it requires $g_H\gtrsim 2\times 10^{-2}$ and $v_\Phi\gtrsim 30$~TeV.


\section{Summary and Conclusion}
\label{sec:summary}

The G2HDM is a novel two Higgs doublet model with a stable DM candidate
protected by an accidental discrete symmetry ($h$-parity) without the need of imposing
it by hand as in the IHDM. After $SU(2)_H$ symmetry breaking, the symmetry remains intact and 
one can find three electrically neutral potential DM candidates with odd $h$-parity: 
the lightest dark complex scalar $D$, heavy neutrino $\nu^H$,
and the $SU(2)_H$ gauge boson $W^{\prime (p,m)}$.  Though these three
candidates are all interesting, we focus this paper on the most
popular one, the new scalar DM $D$, which is complex and hence differ from the DM in IHDM.
Unlike IHDM, the mixing between $\mathcal{Z}$-odd scalars adds a touch of complexity
since DM in G2HDM not only comes from the inert doublet but may also be $SU(2)_H$
Goldstone-like and triplet-like.  We took the
dominant composition ($f_j>2/3$ with $j=H_2,\Delta_p,G^p$) as a criteria to classify them but
the mixture between them can be simply inferred.  In this paper, we have
discussed these three types individually with two assumptions: that all the
new non-SM heavy fermions are heavy enough to have mostly negligible
contributions and that DM were thermally produced before the freeze-out
temperature.  We have comprehensively shown their detectability and exclusions
by the current \textbf{SGSC} and DM constraints (mainly \textbf{RD+DD}). 

Because the DM candidate is chosen to be a complex scalar in G2HDM, 
the DM phenomenology becomes very rich since it has captured 
both features of the Higgs-portal and vector-portal 
DM models discussed in the literature.

For the inert doublet-like DM, we found some interesting features. First,
the main difference between the inert doublet DM in IHDM and G2HDM is that in
IHDM there is in general 
a mass splitting between the scalar $S$ and pseudoscalar $P$ components of $H^0_2$,
while in G2HDM  they are completely degenerate and combined into one single complex field $H^0_2=S+iP$. 
Recall that in IHDM there is only $ZSP$ derivative coupling but no $ZSS$ and $ZPP$ derivative couplings.
As long as the mass splitting between $S$ and $P$ 
remains larger than the exchange energy between DM and nucleons in the direct
detection experiments, the interactions mediated by the $Z$ gauge boson are
suppressed in IHDM. 
Since this splitting does not exist in G2HDM, such interactions are
unsuppressed and they can bring the spin independent cross section up to
$\sim10^{-38}$~cm$^2$, which is significantly above the XENON1T 95\% C.L.  limit
for $m_D \gtrsim$ 10~GeV and above CRESST-III result for $m_D \gtrsim$ 2~GeV
(Fig.~\ref{fig:doublet} right panel).
On the other hand, for $m_D\lesssim 10\gev$, the DM is over
abundant because of on-shell annihilation channels in $c \bar c$ and $\tau^+ \tau^-$
(Fig.~\ref{fig:doublet} left panel).
Hence, we conclude that the inert doublet-like DM can be completely excluded
by \textbf{SGSC+RD+DD} constraints.

Next, a $SU(2)_H$ triplet scalar like DM was discussed.  Since the composition
$f_{H_2}$ has to be tiny in order to avoid the tension with DM {\bf DD}, the
triplet-like DM can mostly mix with the Goldstone boson $G^p$.  There is no
$Z$-resonance region in the triplet-like DM for DM annihilation  and the
parameter space is more or less consistent with Higgs portal DM.
However, \textbf{DD} is still the most stringent constraint comparing with
\textbf{ID} and collider constraints.  The allowed DM mass by
\textbf{SGSC+RD+DD} is required to be heavier than $m_D\gtrsim 300\gev$ 
(Fig.~\ref{fig:triplet} right panel).
Despite weaker constraints coming from \textbf{ID} 
(Fig.~\ref{fig:TLID} left panel) and collider searches, it
might be possible to detect the heavy DM mass region by the future CTA and 100~TeV
colliders even if a DM signal is not found at direct detection
experiments before hitting the neutrino floor. 
As shown by the blue solid boxes in the right panel of
Fig.~\ref{fig:TLID}, the allowed triplet-like DM mass consistent with \textbf{SGSC+RD+DD+ID}
is $\gtrsim$ 300 GeV.

For the last case of the Goldstone-like DM, we found that it is not
possible to obtain a pure Goldstone-like DM. The non-tachyonic DM
condition and EWPT constraints prohibit the composition $f_{G^p}>0.75$ 
(Fig.~\ref{fig:vD_vP_fgp}), unless one would like to move to a more fine-tuned region of parameter space. 
Thus there is a significant component coming from the triplet in the Goldstone-like DM.
Because of the $P$-wave suppression of the $Z$ and $Z^\prime$ exchange
in the dominated channels of $b \bar b$ and $W^+W^-$, 
the annihilation cross section happens to be smaller than for the triplet-like case
and lesser points within the PLANCK relic density measurement (Fig.~\ref{fig:goldstone} left panel). 
Furthermore, XENON1T measurement excludes almost
all the points with appropriate relic density, except
for those with a particular value of isospin violation ($f_n/f_p\approx
- 1.86$) where the sensitivity at XENON1T is reduced.
Therefore, only a small region of orange boxes in the right panel of Fig.~\ref{fig:goldstone}
with $m_D$ in the range of 150 $\sim$  600~GeV can pass all the \textbf{SGSC+DD} 
constraints implemented in this work. 
For \textbf{ID}, the annihilation cross section at the present time for the Goldstone-like DM 
is typically smaller than
the limit
from Fermi gamma-ray constraints
(Fig.~\ref{fig:GLID} left panel).
With significant ISV, only the Goldstone-like DM with a mass in the window of 
150 $\sim$ 600 GeV can be consistent with \textbf{SGSC+RD+DD+ID}, as 
given by the blue solid boxes in the right panel of Fig.~\ref{fig:GLID}.

We also presented the impact of DM constraints on the G2HDM parameter
space in Figs.~\ref{fig:para_lam} and \ref{fig:para_gauge} for the triplet-like DM.
In this case, we found that the following 
parameters $\lambda_\Phi$, $\lambda_{H\Delta}$, $\lambda_{\Phi\Delta}$, 
$g_H$, and $v_\Phi$ are significantly constrained by DM constraints, mainly \textbf{RD+DD},
while the four parameters $g_X$, $\lambda_\Delta$, $v_\Delta$, 
and $M_{H\Delta}$ remains more or less the same as given by the \textbf{SGSC}.
It is interesting to note that the \textbf{SGSC} constraints on 
$g_H$ and $v_\Phi$ as studied in~\cite{Arhrib:2018sbz,Huang:2019obt} 
are now further constrained by \textbf{RD+DD}. 
We note that the lower limit of $g_H>7.09\times 10^{-3}$ for $v_\Phi<100\tev$
is reachable by the future linear (lepton-antilepton) and 100~TeV hadron
colliders.  

Before closing, we would like to make a few comments. 
Originally the $SU(2)_H$ triplet field $\Delta_H$ was introduced to give mass 
to the charged Higgs (Eq.~\eqref{chargedHiggsmass}) in~\cite{Huang:2015wts} where the two 
parameters $\lambda^\prime_H$ and $\lambda^\prime_{H\Phi}$ were missing. 
With these two extra parameters included, the triplet field $\Delta_H$ is no longer 
mandatory. We note however that the triplet field $\Delta_H$ can give rise to 
a non-singular 't Hooft-Polyakov monopole for the hidden $SU(2)_H$ which can play the role as 
DM as studied in~\cite{Baek:2013dwa}~\footnote{We thank P. Ko for bringing this reference to our attention.}. 
Nevertheless, one can have a minimal G2HDM without the triplet field. 
Then the DM $D$ in this minimal model would be just mixture of the inert Higgs $H_2^{0 *}$ 
and the Goldstone field $G^p_H$. From the analysis in this work, we know that this DM scenario 
must be highly fine-tuned in the parameter space due to \textbf{SGSC+RD+DD}.  
A more interesting alternative DM candidate in this minimal G2HDM is the $W^{\prime (p,m)}$,
which certainly deserves a separate study.
Finally, whether the accidental discrete symmetry of $h$-parity, 
identified here in the renormalizable Lagrangian 
for classification of all particles in G2HDM, 
has a deeper origin remains to be explored in the future.

\vfill

\section*{Acknowledgments}
We would like to thank Dr. Wei-Chih Huang for useful  comments and discussions.
TCY would like to thank his host Professor Tri-Nang Pham and the hospitality at 
CPhT of Ecole Polytechnique where progress of the final phase of this work was made.
This work was supported in part by the Ministry of Science and Technology
(MoST) of Taiwan under Grant Nos.\ 
107-2119-M-001-033, 
108-2112-M-001-018 (TCY),
107-2811-M-001-027, 
108-2811-M-001-550 (RR),
and 105-2122-M-003-010-MY3 (CRC).
Y.-L.~S.\ Tsai was funded in part by the Chinese Academy of Sciences Taiwan
Young Talent Programme under Grant No. 2018TW2JA0005.

\newpage

\appendix

\section{Feynman Rules}
\label{sec:appendix}

Here we list the relevant couplings to the DM
analysis in various processes discussed in the text. We use the conventional
notations $g$ and $g'$ to denote the SM $SU(2)_{L}$ and $U(1)_{Y}$
couplings respectively. The $c_W$ and $s_W$ denote the cosine and sine of the
Weinberg angle. The gauge couplings for $SU(2)_H$ and $U(1)_X$ are denoted 
by $g_H$ and $g_X$ respectively.
In addition, for the scalar-scalar-gauge derivative vertices, 
we adopt the convention that all momenta are incoming.

\subsection*{4-point Contact Interaction}
\begin{minipage}{0.3\textwidth}
      \includegraphics[page=5]{feynmp_standalone.pdf}
\end{minipage}
\begin{minipage}{0.69\textwidth}
\begin{flalign}
\:\: = i \frac{e^{2} (\mathcal{O}^{D}_{22})^{2}}{2 s^{2}_{W}} g_{\mu \nu}&&
\label{eq:g_dds_wpwm}
\end{flalign}
\vspace{0.0mm} 
\end{minipage}

\subsection*{Dominant Couplings for Inert Doublet-like DM}

\begin{minipage}{0.3\textwidth}
      \includegraphics[page=6]{feynmp_standalone.pdf}
\end{minipage}
\begin{minipage}{0.69\textwidth}
	\centering
\begin{flalign}
\:\: \approx i \left[\frac{g  c_{W}}{2} + 
\frac{g^{\prime} s_{W}}{2} \right] (\mathcal{O}^{D}_{22})^{2} \mathcal{O}^{G}_{11} (p_{D^{*}} -
p_{D})_{\mu} &&
\label{eq:g_d_ddz1}
\end{flalign}
\vspace{0.0mm} 
\end{minipage}

\begin{minipage}{0.3\textwidth}
      \includegraphics[page=7]{feynmp_standalone.pdf}
\end{minipage}
\begin{minipage}{0.69\textwidth}
\begin{flalign}
&  \approx i \left[\frac{g_{H}}{2}  \mathcal{O}^{G}_{2j} - g_{X}  \mathcal{O}^{G}_{3j}\right] (\mathcal{O}^{D}_{22})^{2} (p_{D^{*}} -
p_{D})_{\mu}\; ,\;\; j = 2, 3 &
\label{eq:g_d_ddz2}
\end{flalign}
\vspace{0.0mm} 
\end{minipage}

\begin{minipage}{0.3\textwidth}
      \includegraphics[page=8]{feynmp_standalone.pdf}
\end{minipage}
\begin{minipage}{0.69\textwidth}
\begin{flalign}
& \approx i \left[-2 \lambda_{H} v \mathcal{O}_{1j} -  \lambda _{H \Phi } v_{\Phi}
	\mathcal{O}_{2j} + \lambda _{H \Delta } v_{\Delta } \mathcal{O}_{3j}
\right] (\mathcal{O}^{D}_{22})^{2} &
\label{eq:g_d_ddh}
\end{flalign}
\vspace{0.0mm} 
\end{minipage}

\subsection*{Dominant Couplings for Triplet-like DM}

\begin{minipage}{0.3\textwidth}
      \includegraphics[page=9]{feynmp_standalone.pdf}
\end{minipage}
\begin{minipage}{0.69\textwidth}
\begin{flalign}
& \approx i g_{H} (\mathcal{O}^{D}_{32})^{2} \mathcal{O}^{G}_{2j} (p_{D^{*}} -
 p_{D})_{\mu} &
\label{eq:g_T_ddz}
\end{flalign}
\vspace{0.0mm} 
\end{minipage}

\begin{minipage}{0.3\textwidth}
      \includegraphics[page=10]{feynmp_standalone.pdf}
\end{minipage}
\begin{minipage}{0.69\textwidth}
\begin{flalign}
& \approx i \left[-\lambda_{H \Delta} v  \mathcal{O}_{1j} -
\lambda _{\Phi \Delta } v_{\Phi }  \mathcal{O}_{2j} + 2 \lambda _{\Delta }
v_{\Delta }  \mathcal{O}_{3j} \right](\mathcal{O}^{D}_{32})^{2} &
\label{eq:g_T_ddh}
\end{flalign}
\vspace{0.0mm} 
\end{minipage}

\subsection*{Dominant Couplings for Goldstone boson-like DM}

\begin{minipage}{0.3\textwidth}
      \includegraphics[page=11]{feynmp_standalone.pdf}
\end{minipage}
\begin{minipage}{0.69\textwidth}
\begin{flalign}
& \approx i \left[ \frac{g_{H}}{2} \mathcal{O}^{G}_{2j} + g_{X}
\mathcal{O}^{G}_{3j}\right] (\mathcal{O}^{D}_{12})^{2} (p_{D^{*}} -
p_{D})_{\mu} &
\label{eq:g_G_ddz}
\end{flalign}
\vspace{0.0mm} 
\end{minipage}

\begin{minipage}{0.3\textwidth}
      \includegraphics[page=12]{feynmp_standalone.pdf}
\end{minipage}
\begin{minipage}{0.69\textwidth}
\begin{flalign}
& \approx i \left[-(\lambda_{H\Phi} + \lambda^{\prime}_{H\Phi}) \mathcal{O}_{1j}  v  -
2\lambda_{\Phi}  \mathcal{O}_{2j}  v_{\Phi} + \lambda_{\Phi \Delta}
\mathcal{O}_{3j}  v_{\Delta} \right](\mathcal{O}^{D}_{12})^{2} &
\label{eq:g_G_ddh}
\end{flalign}
\end{minipage}


\end{document}